\documentclass{aa}
\usepackage[varg]{txfonts}

\usepackage{graphicx}
\usepackage{txfonts}

\usepackage{natbib,twoopt}
\usepackage[breaklinks=true]{hyperref} %% to avoid \citeads line fills
\bibpunct{(}{)}{;}{a}{}{,}             %% natbib format for A&A and ApJ
\makeatletter
  \newcommandtwoopt{\citeads}[3][][]{\href{http://adsabs.harvard.edu/abs/#3}%
    {\def\hyper@linkstart##1##2{}%
     \let\hyper@linkend\@empty\citealp[#1][#2]{#3}}}
  \newcommandtwoopt{\citepads}[3][][]{\href{http://adsabs.harvard.edu/abs/#3}%
    {\def\hyper@linkstart##1##2{}%
     \let\hyper@linkend\@empty\citep[#1][#2]{#3}}}
  \newcommandtwoopt{\citetads}[3][][]{\href{http://adsabs.harvard.edu/abs/#3}%
    {\def\hyper@linkstart##1##2{}%
     \let\hyper@linkend\@empty\citet[#1][#2]{#3}}}
  \newcommandtwoopt{\citeyearads}[3][][]%
    {\href{http://adsabs.harvard.edu/abs/#3}
    {\def\hyper@linkstart##1##2{}%
     \let\hyper@linkend\@empty\citeyear[#1][#2]{#3}}}
\makeatother

\newcommand{\Mrate}{\dot{M}}
\newcommand{\Mdot}{\mbox{\,$\rm M_{\odot}$}}        % solar mass
\newcommand{\Ldot}{\mbox{\,$\rm L_{\odot}$}  }        % solar luminosity
\newcommand{\kms}{\,km s$^{-1}$}   			% kms-1
\newcommand{\aov}{$\alpha_{\rm ov}$}		% overshoot alpha
\providecommand{\keywords}[1]{\textbf{\textit{Keywords--}} #1}
\newcommand{\vsini}{$\varv \sin i$}

\begin{document}

\titlerunning{<Massive star evolution>}
\title{Massive star evolution : rotation, winds, and \\overshooting vectors in the Mass-Luminosity plane}
\subtitle{I. A calibrated grid of rotating single star models\thanks{ Evolutionary tracks are available in electronic form at the CDS via \url{http://cdsweb.u-strasbg.fr/cgi-bin/qcat?J/A+A/} or via either authors homepage e.g. \url{http://193.63.77.2:8383/armaghobservatoryplanetarium/published/erin_higgins.php} , or by contacting either author.}\\ }

\author{Erin R. Higgins\inst{\ref{inst1}}\inst{\ref{inst2}}\inst{\ref{inst3}}\inst{\ref{inst4}}\and Jorick S. Vink\inst{\ref{inst1}}\inst{\ref{inst4}}}

\institute{Armagh Observatory and Planetarium, College Hill, Armagh BT61 9DG, N. Ireland\label{inst1} \and Queen's University of Belfast, Belfast BT7 1NN, N. Ireland\label{inst2} \and Dublin Institute for Advanced Studies, 31 Fitzwilliam Place, Dublin, Ireland\label{inst3} \and Kavli Institute for Theoretical Physics, University of California, Santa Barbara, CA 93106, USA\label{inst4} \\email{: erin.higgins@armagh.ac.uk; jorick.vink@armagh.ac.uk}}

\date{Received 22 August 2018 / Accepted 29 November 2018}
\abstract
{Massive star evolution is dominated by various physical effects,  including mass loss, overshooting, and rotation, but the prescriptions of their effects are poorly constrained and even affect our understanding of the main sequence. }{We aim to constrain massive star evolution models using the unique test-bed eclipsing binary HD\,166734 with new grids of MESA stellar evolution models, adopting calibrated prescriptions of overshooting, mass loss, and rotation.}{We introduce a novel tool, called the mass-luminosity plane or $M-L$ plane, as an equivalent to the traditional HR diagram, utilising it to reproduce the test-bed binary HD\,166734 with newly calibrated MESA stellar evolution models for single stars.} {We can only reproduce the Galactic binary system with an enhanced amount of core overshooting (\aov $=$ 0.5), mass loss, and rotational mixing. We can utilise the gradient in the $M-L$ plane to constrain the amount of mass loss to 0.5 - 1.5 times the standard prescription test-bed, and we can exclude extreme reduction or multiplication factors.  The extent of the vectors in the $M-L$ plane leads us to conclude that the amount of core overshooting is larger than is normally adopted in contemporary massive star evolution models. We furthermore conclude that rotational mixing is mandatory to obtain the correct nitrogen abundance ratios between the primary and secondary components (3:1) in our test-bed binary system.} {Our calibrated grid of models, alongside our new $M-L$ plane approach, present the possibility of a widened main sequence due to an increased demand for core overshooting. The increased amount of core overshooting is not only needed to explain the extended main sequence, but the enhanced overshooting is also needed to explain the location of the upper-luminosity limit of the red supergiants. Finally, the increased amount of core overshooting has  -- via the compactness parameter -- implications for supernova explodability.}

\maketitle
\section{Introduction}
Massive stars with an initial mass above 8\Mdot\ have a diversity of possible evolutionary channels that are dictated by the dominant processes acting on their structure. The extent of these dependencies are variant with mass, metallicity, and multiplicity. Stellar winds have a significant impact on the evolution of O-type stars throughout their lives, leading to evolutionary phases involving Luminous Blue Variables (LBV) and Wolf-Rayet (WR) stars. It is also an important factor for dictating their final masses and determining whether a neutron star or black hole is formed in the final stage of evolution, as extensively reviewed by \cite{Chiosi86}. 

On the main sequence (MS), mass loss via stellar winds has the greatest impact at the highest mass ranges. Above $\simeq$60\Mdot\ mass loss completely dominates the evolution of O-type stars \citep[e.g.][]{VG12, V15}, whilst in the range 30\Mdot $\textless$ $M$ $\textless$ 60\Mdot\ mass loss is one of the important ingredients \citep[e.g.][]{Langer12, Groh14}. 
At lower masses, (i.e. below $\sim$ 30\Mdot) the evolution is thought to be heavily influenced by rotation \citep[e.g.][]{MM00}. Over the explored mass range within this paper, i.e. 8 - 60 \Mdot, we will consider the effects of mass loss and rotation, as well as convective overshooting, which may all play a role in the evolution of these stars.

The extension of the convective core by overshooting is a key structural feature which increases the amount of hydrogen (H) dredged into the core, replenishing its supply, thereby extending the MS lifetime. The parameter \aov~ which we explore in this study corresponds to the fraction of the pressure scale height H$_\text{p}$ by which particles continue to travel a distance $l_\text{ov}$ beyond the convective core boundary. This form of mixing has been explored for decades, with few constraints on its size (\aov) in the high-mass range. It has been argued as essential for reproducing observations, although evidence is lacking for dependencies such as mass \citep{ClaretTorres}. Another process that may potentially affect stellar evolution is the presence of a magnetic field, however, \cite{wade12} showed the fraction of magnetic O stars to be just on the order of 7\%.

Massive star evolution models are currently not able to fully reproduce observations, even the MS \citep[e.g.][]{Vink10, Markova}, with many physical processes such as rotation and overshooting yet to be fully understood. The MS width and dependencies remain unresolved, while this stage represents 90\% of the overall lifetime and sets the stage for later evolutionary phases. 

Efforts have been made to map out the evolution of massive stars with systematic grids \citep[e.g.][]{Bonn11, Gen12} using detailed predictions of chemical abundances, rotation rates, and fundamental parameters such as mass, luminosity, and effective temperature. These models have subsequently been compared to observations for predicting evolutionary stages and characteristics, though due to limitations in both key observations and accurately modelling key physical processes, many assumptions remain, including the amount of core overshooting (\aov) that is thought appropriate. 

\cite{Martins13} explored a diversity of evolutionary codes \citep[e.g.][]{Gen12, Chieffi, Bertelli} in which the implementation of input physics was surveyed, allowing code applicability to be tested, however linear comparisons of physical treatments cannot be drawn due to the variety of prescriptions in different codes. It is clear that all stellar evolution models have a degree of uncertainty, yet to establish a clear comparison between codes, it would be beneficial to examine physical implementations with one and the same code. Therefore, in this work we aim to compute massive star models with both new and existing prescriptions using the same evolutionary code, Modules for Experiments in Stellar Astrophysics  \citep[MESA; e.g.][]{Pax11} given its high flexibility and code capabilities, enabling ample comparisons of several key physical processes. Such exploration offers the opportunity for calibrating models with respect to observations.

Due to the variety of possible prescriptions in each code, the evolution of massive O-type stars so far remains model dependent, leaving the MS lifetime ambiguous particularly due to the absence of evidence for objects after the terminal age main sequence (TAMS). 
Cool B supergiants, the descendants of O-type stars, are less understood, and have yet to be confirmed as core hydrogen or helium burning objects, \citep{Vink10}. As O-type stars spend the majority of their lifetime on the MS, we would expect a scarcity of B supergiants if they are indeed post-MS objects. However, we observe a (too) large number of these stars \citep[e.g.][]{Garmany}, raising the possibility that these objects are MS, core H-burning stars. The existence of a large number of slow-rotating B supergiants however, (with \vsini  $\lesssim$ 50 \kms) is suggestive of an evolved star that has completed the MS phase and been spun down. 

\cite{Vink10} also considered the possibility for bi-stability braking (BSB) as the mechanism by which B supergiants lose their angular momentum \citep[see also][]{Zsolt17}. If we consider that B supergiants may not represent the end of the H-burning phase, this could allow for a wider MS, hypothesised by \cite{Vink10}. This would result in a demand for additional mixing of H in the core, which may be fulfilled by increased overshooting. \cite{Vink10} addressed that a higher value of \aov~ would result in a lower critical mass at which BSB would be efficient. Test models show that BSB occurs in present models with \aov$=$0.335 above a critical mass of 35\Mdot\ in the Large Magellanic Cloud (LMC), yet the critical mass drops to 20\Mdot ~for the same metallicity with an increase of overshooting to \aov$=$0.5 \citep{Vink10}.

The determination of \aov\ for massive stars has been challenging without the aid of astroseismological data for the most massive stars, leading to an array of prescriptions such as the correlation between \vsini~ and log \textit{g} \citep{Bonn11}. Many other estimations of \aov~ have been adapted in stellar evolutionary models leading to a wide variety of potential stellar ages, MS lifetimes, and final products \citep{Martins13}.
One of the most straightforward approaches would be to derive it simply from the MS width, which might potentially be possible from the Galactic Hertzsprung-Russell diagram (HRD) in \cite{Castro14}. However, the Galactic data from this study might be biased compared with the unbiased LMC data from the VLT-Flames Surveys of massive stars \cite{Evans05, Evans11} as they do not show a gap between O and B supergiants \citep{Vink10}, thereby suggesting a more extended MS.

In this paper, we attempt to constrain the dominant parameters effecting massive star evolution. For this purpose, we present a grid of evolutionary models for two extreme values of \aov $=$ 0.1 and 0.5 to illustrate both lower adopted values and enhanced overshooting, with varying initial masses, rotation rates, and mass-loss rates, thereby highlighting the sensitivity of stellar models in terms of mixing and mass loss. We introduce the Mass-Luminosity Plane as an alternative to the HRD to study the key ingredients in massive star evolution on the MS (see Fig.\,\ref{Cartoon}). Whilst the fundamental stellar parameters of mass and luminosity have been plotted logarithmically by \cite{1983M} for example, our version of the plot highlights the independent effects of rotation, overshooting and mass loss on stellar evolution through vectors, with inverted mass on the x-axis providing a useful comparison to the tracks in the HRD.

\citet{WV10} presented an overview of the methods of mass determination for O stars,  including the 'mass discrepancy' often seen between the evolutionary masses and spectroscopic masses. The method of comparing the positions of stars in the HRD with theoretical evolution models (evolutionary masses) has often led to predictions which are systematically higher than the masses derived through stellar spectroscopy (spectroscopic masses) \citep[e.g.][]{Herrero}. However, when O stars are found in binary systems, their dynamics can present a model independent mass determination (dynamical masses). Evolutionary masses can present discrepancies amongst themselves when using various theoretical models \citep[e.g.][]{Gen12, Bonn11} with differing implementations of rotation, convection and mass loss. Although this is not the widely discussed ‘mass discrepancy’ problem, it does highlight the necessity of calibrating stellar evolution models to minimise further discrepancies with spectroscopic and dynamical masses \citep[see e.g.][]{Markova}. In the case where dynamical masses agree with the spectroscopic masses, we can have faith in the spectroscopic result,  thus allowing for calibration of theoretical evolution models. Similar work has been completed by \citet{South, Pav, Tka} for detached eclipsing binaries, however these works utilised lower mass stars (up to $\sim$15\Mdot) which did not incorporate the interacting effects of mass loss, overshooting and rotation as we do in this study.

We use constraints relative to \aov\ and $\dot{M}$\ to investigate the possible evolutionary paths of a high mass, detached binary, HD\,166734, modelled in this work as a test-bed for single star evolution. As previously mentioned, dominant processes take effect at varying mass ranges, yet with dynamical masses of 39.5\Mdot\ and 33.5\Mdot\, for the primary and secondary, respectively, for HD\,166734 we may probe the effects of these processes as they interact and overlap. As the spectroscopic masses adeptly agree with the dynamical masses for HD\,166734 \citep{Mahy2017}, this system provides a unique opportunity to constrain -- and effectively correct -- stellar evolution models, whilst for the general case of single massive stars we cannot currently tell if there are issues with spectroscopic masses, leading to mass discrepancies \citep{Markova}. 

We present a method of producing a calibrated grid of models with an analysis of HD\,166734 in section \ref{calibration} and a calibration of mixing processes in section \ref{3}. We explore a new tool for comparing observations with models in the Mass-Luminosity Plane in section \ref{4} and we provide our final results for HD\,166734 in section \ref{5}. We present our grid of models alongside a sample of Galactic O-stars in section \ref{6}, and provide further results in Appendix \ref{gridapp}. Finally, we highlight our conclusions in section \ref{7}; the remaining full grid plots are available in Appendix \ref{gridapp}.

\section{Methodology}
\subsection{MESA : Treatment of convection, mass loss, and rotation}
\label{calibration}
A set of evolutionary models was calculated for massive MS stars with the 1D, stellar evolution code MESA, for example \citet{Pax11}, as a comparison for both the primary and secondary of HD\,166734 (see Sect.\,\ref{HD166734}). The extensive capabilities of this code provide a diverse range of available alterations, enabling the user to compare implementations of physical processes with other code treatments. In this paper, we examine the effects of mass loss, convective overshooting, and rotational mixing in terms of fundamental observables such as luminosity, mass, and surface abundances. These models were completed from zero-age main-sequence (ZAMS) to core collapse, unless convergence problems arose in which computations were concluded earlier. We adopt the default metallicity in MESA of Z = 0.02 with the chemical mixture from \citet{Grev98} to provide direct comparisons with chemical abundances in Galactic observations. 

Convection is treated by the mixing length theory where {$\alpha_{\rm MLT}$} $=$ 1.5, with a semi-convection efficiency parameter of {$\alpha_\text{semi}$} $=$ 1. The convective core boundary is defined by the Ledoux criterion\footnote{ The Ledoux criterion is denoted by $\nabla_{rad}$  $<$ $\nabla_{ad}$  $+$ $\frac{\phi}{\delta}$ $\nabla_{\mu}$ , but in chemically homogeneous layers where $\nabla_{rad}$ = $\nabla_{ad}$ the Schwarzschild criterion is effective. }, in which overshooting succeeds convective mixing at the core boundary, increasing the temperature gradient $\nabla_{T}$ by implementing a thermal gradient $\nabla_{rad}$ \citep[e.g.][]{MIST}. This method of extending the core is denoted as step-overshooting, which enhances the core by a factor \aov \ of the pressure scale height $H_\text{p}$. Experiments in the dependencies of this parameter are completed in the following sections.

We then compared our grid with treatments of \aov~ and rotational mixing from \cite{Bonn11} and \cite{Gen12} grids since these are used extensively in the community. \cite{Bonn11} presented a calibration of the overshooting parameter by comparing the TAMS of 16\Mdot~ models with observations from the FLAMES survey \citep{Evans08}, suggesting a TAMS at log \textit{g} $=$ 3.2, since this value represents a drop in \vsini~ beyond which a large number of slow-rotating B supergiants are located, assumed to be post-MS objects. The model with \aov~$=$0.335 corresponded to the log \textit{g} $=$ 3.2 and has since been used as a static parameter in models to compare against observations of a wide mass range. A lower value of \aov$=$0.1 is applied for models presented by \cite{Gen12}, since calibration was completed with lower mass stars of $\sim$1.7-2\Mdot~ where convective mixing plays a dominant role compared to that of rotational mixing. Hence this allowed for a linear calibration of convective overshooting without accounting for the more sophisticated treatment of rotational mixing as prescribed in the GENEC code. We adopt step-overshooting for H-burning phases only, as we aim to better understand the MS width.

Mass-loss rates are adopted from \citet{Vink01} accounting for metallicity dependencies and the occurrence of the bi-stability jump, an increase of mass loss at 21kK causing effects in the evolution, seen in the HRD. We tested various factors of this mass-loss regime to determine the possibility of extreme rates. We hence explored a range of multiplication factors of \cite{Vink01} mass-loss rates from 0.1 to 10 times the standard prescription. We later applied rotation in our models through a fully diffusive approach with appropriate instabilities such as the Eddington-Sweet circulation, dynamical and secular shear instabilities. We also considered the effects of an internal magnetic field by a Spruit-Taylor dynamo, although we found that this had inconsequential effects on our results. The calibration of our single star models are relevant for evolutionary codes which implement rotational mixing in a similar way. If this process is treated as physically different in another code, then the results would differ quantitatively, but qualitatively have the same behaviour.

\begin{table}
\caption{\label{grid}Calibrated grid of stellar evolutionary models.}
\centering
\begin{tabular}{l|c}
\hline\hline
$M_\text{initial}$ [\Mdot] & 8, 12, 16, 20, 25, 30, 35, 40, 45, 50, 55, 60\\
$\varv_\text{initial}$ [\kms] & 0, 100, 200, 300, 400, 500\\
\aov & 0.1, 0.5\\
\hline
\end{tabular}
\end{table}

A systematic grid of models was calculated for comparison with a larger sample, including new prescriptions discussed in Sections \ref{3} and \ref{4}. Table \ref{grid} shows the range of masses, rotation rates, and overshooting values for which we compose our grid. We choose masses representative for the O-star and early B-star range, with a variety of rotation rates up to break-up speed, and extreme values for \aov~  to explore the extent of extra mixing. We evolved each model to core collapse, unless convergence problems highlighted unlikely solutions. For this purpose, \cite{Vink01} provided the relevant mass-loss prescription, with a factor of unity for all models in the first instance. 

\subsection{The detached, eclipsing binary HD\,166734 : A test-bed for massive star evolution}\label{HD166734}

The eclipsing massive binary HD\, 166734 (see Table \ref{HD166}) provides a unique opportunity to improve physics in stellar evolution models as \cite{Mahy2017} were able to determine the individual stellar parameters, including their exact positions in the HRD and their dynamical real masses directly. 
As these dynamical masses were found to be in excellent agreement with their spectroscopic masses, these two stars of this massive binary system, enable us to calibrate and correct the evolutionary masses, thereby constraining the relevant physics in the upper HRD for stars above 30-40\Mdot. 
Observations of high-mass eclipsing binaries are sparse, and even more extreme for detached, non-interacting stars that may be treated as evolved single stars. Since observations of massive single stars may sometimes highlight discrepancies between spectroscopic and evolutionary masses. In this case, we have an ideal opportunity because the dynamical masses are in agreement with spectroscopic masses, providing a tool for calibrating evolutionary masses and thus evolutionary paths of stars that are massive enough for the physics to be heavily influenced, if not dominated by mass loss via stellar winds.

Though a large fraction of O stars may be present in a binary or multiple system, observations of eclipsing binaries above ~30\Mdot~ are extremely rare \citep[see e.g.][]{Bonanos, deMink, Pfuhl, Gies}. Hence the stellar parameters derived by \citet{Mahy2017} have provided a unique opportunity to analyse a non-interacting system which can be treated as single stars. The similar values of \vsini~ for both components may at face value be considered of interest in terms of synchronisation, but \cite{Mahy2017} have argued against synchronisation because the rotation speeds are lower than the orbital period. In addition, we note that the \vsini~ values are close to the inferred macro-turbulent values of 65 $\pm$ 10 \kms \citep{Mahy2017} and we therefore urge for caution that the quoted values of \vsini~ are truly the result of rotation \citep[see][]{Sdiaz}. We thus treat the \vsini~ values as upper limits, and we consider the similar values of the two components as merely a coincidence.
\begin{table}
\caption{\label{HD166}HD\,166734 Properties}
\centering
\begin{tabular}{lcc}
\hline\hline
 & Primary & Secondary\\ 
\hline
$T_\text{eff}$ [K] & 32000 $\pm$1000 & 30500 $\pm$1000\\
log(L/\Ldot) & 5.840 $\pm$ 0.092 & 5.732 $\pm$ 0.104\\
$M_\text{dyn}$ [\Mdot] & 39.5 $\pm$ 5.4 & 33.5 $\pm$ 4.6\\
$M_\text{spec}$ [\Mdot] & 37.7 $\pm$ 29.2 & 31.8 $\pm$ 26.6\\
\vsini [\kms] & 95 $\pm$ 10 & 98 $\pm$ 10\\
$\rm [N/H]$  & 8.785 & 8.255\\
\hline
\end{tabular}
\tablefoot{
Fundamental observational properties of HD\,166734, adapted from \cite{Mahy2017}.
}
\end{table}
We utilise this agreement between dynamical and spectroscopic masses, allowing HD\,166734 to be treated as an excellent test-bed for massive star evolution of the most massive O-type stars. \cite{Mahy2017} analysed the system finding a composition of two supergiant O-type stars in an eccentric 34.5-day orbital period.
We recognise that the estimated \vsini\ quantities may be upper limits due to the possibility of macro-turbulence. We can also compare with observed surface N abundances as a secondary assessment of potential rotation rates. 

Comparisons to fixed current-day evolutionary sets of models by \cite{Bonn11} and \cite{Gen12} by \cite{Mahy2017} revealed that both sets of models over-predict the evolutionary masses, whilst the secondary star appeared to be more evolved than the primary.
We consider this latter finding an artefact of the \cite{Mahy2017} approach, and that in reality it is far more likely that both components formed simultaneously. We can therefore use an equal-age assumption in addition to the exact HRD positions and true current day masses to solve the evolutionary mass discrepancy for both components, and at the same time constrain the relevant physics in this mass range.

Our assumption that this binary has evolved from the same initial stage is important for constraints of the MS width and thus for constraining the overshooting parameter and determining the rotation rates and possible evolutionary scenarios. As both stars show limited evidence of an evolved nature, we can exclude extreme events in the past such as eruptive mass loss or binary interactions. \cite{Mahy2017} showed surface nitrogen enrichments with a particle fraction [N/H] ratio of 3:1 between the primary and secondary components respectively. We utilise these abundances as evidence for mixing, as well as constraints for the determination of age.

\section{Mixing and mass loss}\label{3}
\subsection{Envelope stripping and nitrogen enrichment.}

In developing our initial set of models we aim to minimise interacting physical processes. We start with a set of non-rotating stellar evolution models that exclusively employ mass loss and convective overshooting as mixing processes. In the first instance, initial masses were adopted from \citet{Mahy2017} with 56.1 \Mdot ~and 47.4 \Mdot ~for the primary and secondary, respectively; with varying factors of the mass-loss recipe adopted from \citet{Vink01} for a range of convective overshooting parameters \aov. We initially attempted to reproduce characteristics of HD\,166734 by following analysis from \cite{Mahy2017} with parameters taken from \cite{Bonn11} and \cite{Gen12} grids. We found however that these models do not offer solutions in which sufficient N enrichment is reached. We hence employ greater mixing through increased factors of mass loss and overshooting.

In reproducing the properties of HD\,166734, we can constrain the scenarios that display the 3:1 ratio of [N/H] for the primary to secondary by applying a restriction to the model time. As both stars are assumed to be approximately the same age with this ratio of enrichment, we can exclude the vast majority of possible evolutionary scenarios, i.e. those that do not represent these surface chemical enrichments simultaneously. Accordingly, we do not predict the ages of these stars, but we rather allow for constraints such as surface enrichments, rotation rates, and dynamical masses to provide a solution whereby both stars can reproduce the observables concurrently. Analogous to this, isochrones have not been used here as a method of stellar age determination as we have previously highlighted the sensitivity of model dependency on these features, thus leading to a wide range of possible ages. 

Massive stars produce surface He on the MS by the CNO cycle, with a rapid increase in \textsuperscript{14}N by a factor of $\sim$10 at the surface when CN equilibrium is reached. The occurrence of this observational feature has been reviewed widely by \cite{MM87}, finding that increased convective mixing by overshooting has shown to lower the limit for CN equilibrium during the MS. 

\citet{MM87, MM88, MM91} composed grids of evolutionary models based on inputs of mass-loss rates and convective overshooting \aov\ as the sole mechanisms for chemical mixing. The importance of convective overshooting has been stressed in these early publications as \aov\ leads to a range of stellar ages, due to the dependence of T$_{\rm eff}$  at TAMS on \aov, \citep{MM91}. Moreover, the MS luminosity increases by 0.9 dex at the reddest point of the MS when overshooting is accounted for leading to increases in age by factors of 1.5 - 2.7. 

\begin{figure}
\centering
	\includegraphics[width = 9cm]{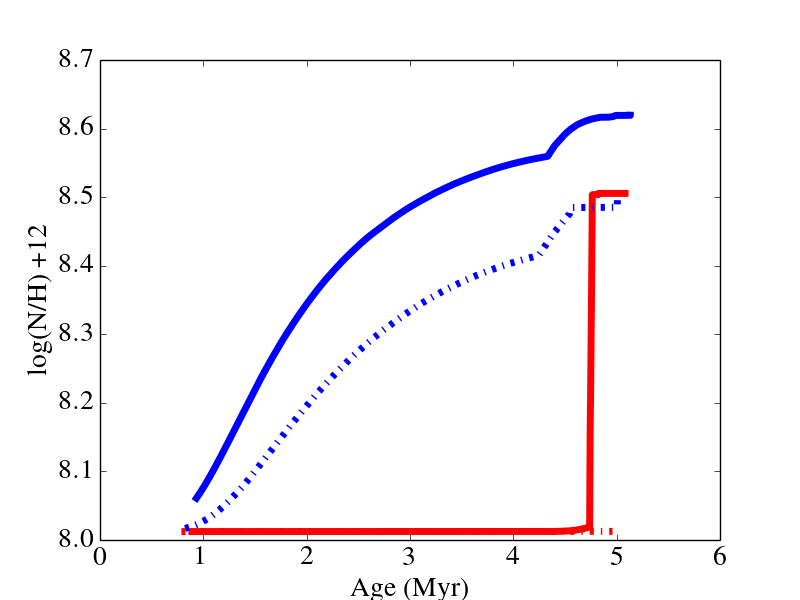}
	\caption{\footnotesize Surface nitrogen abundances as a function of stellar age for extreme values of \aov~ and \Mdot~. The blue lines represent rotating 40\Mdot~ models with an initial rotation rate of 200\kms, \aov=0.1 (dash-dotted), and \aov=0.5 (solid). The red lines show the corresponding non-rotating models for the same mass and values of \aov\ respectively.}
	\label{N/H}
\end{figure}

\cite{MM94} presented grids of massive stars with high mass-loss rates since the evolution of the most massive stars is so heavily reliant on the effect of stellar winds. A factor of two enhancement was applied to their mass-loss prescription from \cite{SCH92} demonstrating the effects on the evolutionary track presented in a HRD. These results hinted at a metallicity dependency on mass-loss rates, though also show envelope stripping with increased mass loss leading to evolutionary phases such as WR types and quasi-chemical homogeneous evolution \citep{MM94}.
When analysing nitrogen enrichments for these models we find that if surface abundances do increase, it is by a sudden step of a factor of ten, representative of CN-equilibrium. This behaviour applies to factors of 1 - 3 of \cite{Vink01} mass-loss rates, and overshooting \aov~ of 0.1 - 0.8. We also note that models with increased overshooting result in earlier enrichment by up to 1Myr, regardless of mass-loss rates. In figure \ref{N/H} we present the nitrogen enrichments for a sample of models of primary and secondary masses. 

We find that chemical mixing of CNO elements by mass loss and overshooting attains CN equilibrium before any intermediate enrichment occurs. This demonstrates that a combination of increased mass loss and overshooting results in envelope stripping, whereby fusion products are extensively exposed at the stellar surface. Since this does not provide a solution for reproducing the observed surface enrichments of HD\,166734, and moreover any observation with intermediate enrichment, we must explore additional, viable mixing processes, such as rotational mixing. 

As we have adopted the method of step-overshooting, physical implications of this may hinder intermediate enrichment since step-overshooting invokes instantaneous homogeneous mixing within the overshooting region, leading to immediate enrichment by a factor of 10 when the envelope is stripped via stellar winds. Therefore, we compared our results with the prescription of exponential overshooting, whereby the length of the scale height is set by a comparable parameter f$_{0}$, but the overshooting region is mixed by a diffusion gradient. 

Nevertheless, these results show similar enrichments, as even though intermediate enrichments may be reached through the overshooting region by altering the diffusion coefficient, elements are not mixed intermittently through the envelope from the overshooting layer. Thus another mixing process capable of mixing the chemical elements from the convective layers through the envelope must be implemented in order to match observed enrichments.

Recent studies of massive star observations \citep[e.g.][]{Brottb, Hunter08b, Maeder00} suggest that surface enrichments of CNO products may or may not be a result of rotational mixing.  Yet, the necessity of rotational mixing has not been stressed with respect to CN equilibrium or observed intermediate enrichments. We therefore tested the effects of rotational mixing as a function of surface enrichment, with a set of rotating models of varied initial rotation rates from 100-500 \kms. 
In this set of models we find that a range of intermediate enrichments occurs, also providing solutions for reproducing the 3:1 nitrogen ratios as seen in HD\,166734, (Fig.\ref{N/H}). The comparison in Fig. \ref{N/H} illustrates that rotational mixing is essential in reproducing observational surface enrichments, unless another not yet considered mechanism is identified, since previous mixing processes provide either too little or too much mixing leading to insignificant enrichment or CN equilibrium.

Fig.\, \ref{N/H} not only demonstrates the necessity of rotational mixing, but also stresses the importance of enhanced overshooting. In the rotating models of Fig.\, \ref{N/H} we see that with an increase in \aov~ from 0.1 to 0.5, we get much larger surface enrichments that may aid our understanding of the unexplained nitrogen enrichments discussed by \citet{Grin17}. As a significant fraction of the sample cannot be explained by rotational mixing alone, extended overshooting may help towards resolving this problem.

\subsection{Rotationally induced mass loss}
While analysing a set of rotating models for HD\,166734 we discovered a problem with respect to interacting processes such as rotation and mass loss, which consequently have a non-linear effect on the mass and luminosity. We find that the initial masses sufficient for reproducing the observed luminosities are excessive when aiming to reach the dynamical masses by the time of observed temperatures or evolutionary phases. 
We therefore calculated a set of lower initial mass models, yet these diminish the luminosity gradient over time so that current data points of 
HD\,166734 remain out of reach. Interpreting an initial mass from the observed luminosity allowed for calculation of a possible mass-loss rate that would enable the current dynamical masses to be reached. 

Following this method, we find a mass-loss rate of log $\dot{M}$ $=$ $-$ 5.17, translating to an increase in the mass-loss rate by approximately a factor of 3. We therefore completed further models with increased mass-loss rates of a factor of two and three. We now reached the dynamical masses; this also led to a significant drop in luminosity, which correlates to a shallow gradient in the $M-L$ plane (see Fig.\, \ref{MLHD}.), suggesting the observed masses and luminosities could not be reproduced simultaneously (see Fig. \ref{Quadrant}). 
The possibility of rotationally enhanced mass loss started with 1D radiation-driven wind models of \cite{FriendAbbott86}, who proposed an equatorially enhanced stellar wind and an increased mass-loss rate due to a lower effective gravity at the equator. This result is also included in many massive star evolution models \citep[see][]{HLW00, Bonn11}. This same implementation is included in the default MESA settings. The mathematical approach in shown in Eq.\,(\ref{mdotboost}), i.e.  

\begin{equation}
\Mrate =\bigg( \frac{\Mrate_{0} }{ \frac{1}{1 - \Omega}}\bigg)^{\xi} ~~~where ~~~\xi = 0.43.
\label{mdotboost}
\end{equation}  

We note that the Geneva group \citep[e.g.][]{MM15,Gen12} employed a slightly different implementation, yet it is based on similar physical principles. Since 1986 there have been many studies of the effects of rotation on radiation-driven wind predictions with several different levels of sophistication, and different results. Recent 2D modelling by \cite{MullerVink} encountered cases of equatorial decreases of the mass-loss rate and surface-averaged total mass-loss rates that are lower than for the 1D case. They therefore challenged the implementation of rotationally enhanced mass-loss in stellar evolution modelling, which is still mostly applied; for example it is the default setting in MESA. 

\begin{figure}
\center
\includegraphics[width = 9cm]{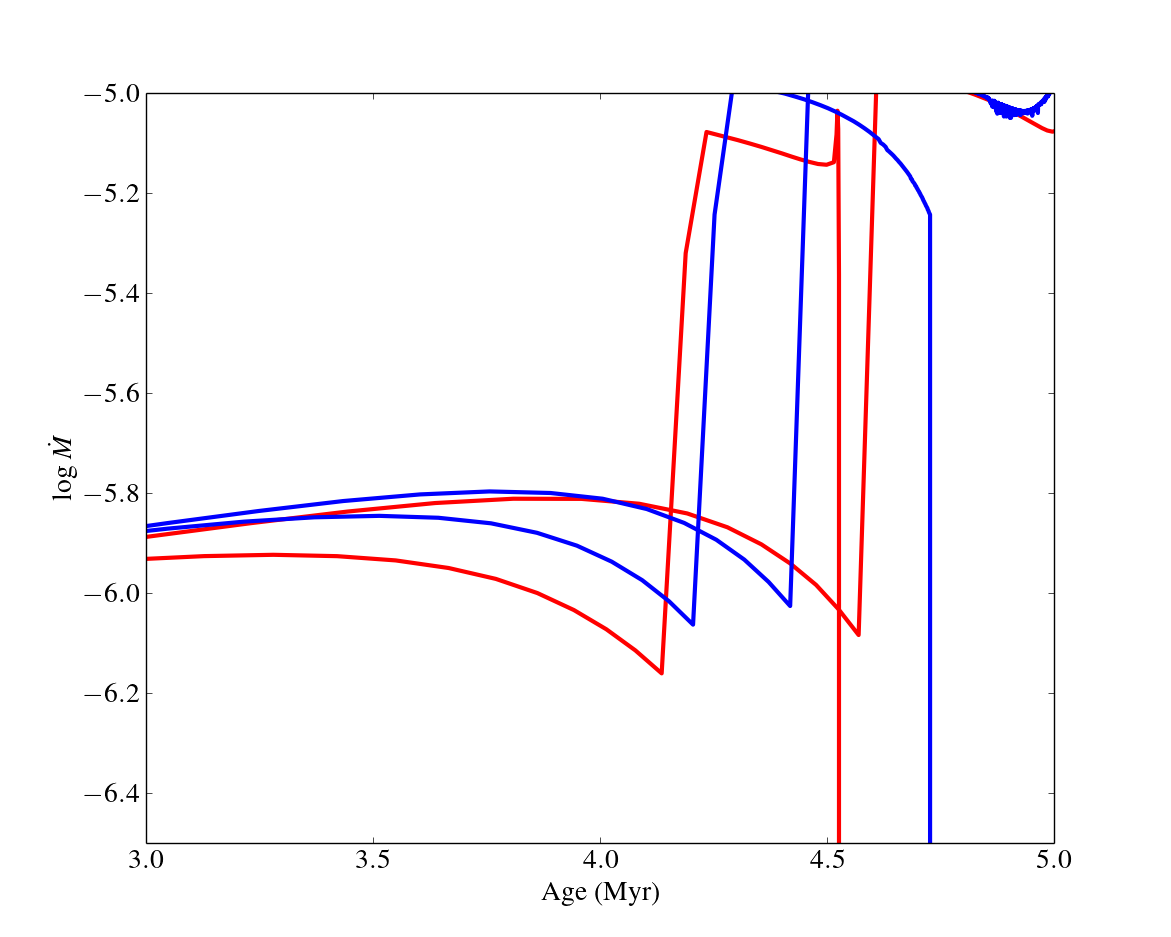}
\caption{\footnotesize Mass-loss rate as a function of stellar age for comparison of rotationally induced mass loss. The red solid lines represent models with default MESA settings of rotationally induced $\dot{M}$ (see Eq.\,(\ref{mdotboost})), with an initial mass of 40\Mdot~ and rotation rates of 100\kms and 300\kms. All other processes have been set to default values to avoid conflict in our analysis. The dashed blue lines show the corresponding models with $\xi$ $=$ 0, for the same mass and rotation rates respectively.}
\label{Mdott}
\end{figure}

Figure \ref{Mdott} highlights the change in initial mass-loss rate due to a change in initial rotation rate from 100 - 300 \kms\, for both a 40\Mdot\, model. As we consider the current enhancement largely as artificial, we explored the difference between disabling rotationally enhanced mass loss (effectively setting $\xi$ $=$ 0) and enabling it using the default setting ($\xi$ $=$ 0.43). We thus calculated a series of models with various initial masses, rotation rates, mass-loss rates and overshooting parameters. 

\section{Mass - Luminosity Plane}\label{4}
When comparing models in Section \ref{3} we find that enhanced mass-loss regimes lead to unrealistic luminosities that are too low to reproduce the observed HD\,166734 luminosities. We also find that an initial mass representative of the observed luminosities is too high to reproduce the much lower dynamical masses with factor unity of \cite{Vink01} mass-loss rates. If we aim to simultaneously reproduce the mass and luminosity, we must explore all possible dependencies of these properties,\begin{equation}
\centering
L = \mu^4 M^\alpha ,
\label{ML}
\end{equation} ~~~~~~where $\alpha$ varies as a function of mass and $\mu$ is the mean molecular weight.\\
The most fundamental characteristics of the evolution of a star are its mass and luminosity. As such, when trying to correlate the theoretical evolution of a star with its observables, these properties are essential. Thanks to analysis of HD\,166734 by \cite{Mahy2017}, we can reliably utilise the luminosities of both stars determined from bolometric magnitudes to calibrate their evolutionary status. This reasoning is also applicable to the masses of HD\,166734, as in this circumstance the dynamical masses agree very well with the derived spectroscopic masses, providing a unique opportunity to constrain the mass-loss rates and physical processes during evolution. 

The mass and luminosity of a star are reliant on age and mass-loss rate, so we reach a diversity of possible evolutionary scenarios with respect to mass-loss rates and \aov. Yet we may constrain these solutions by assuming both objects evolved from the same initial starting point, so we can account for primary and secondary masses to be reached at the same time. 
\begin{figure}
\centering
\includegraphics[width = 9cm]{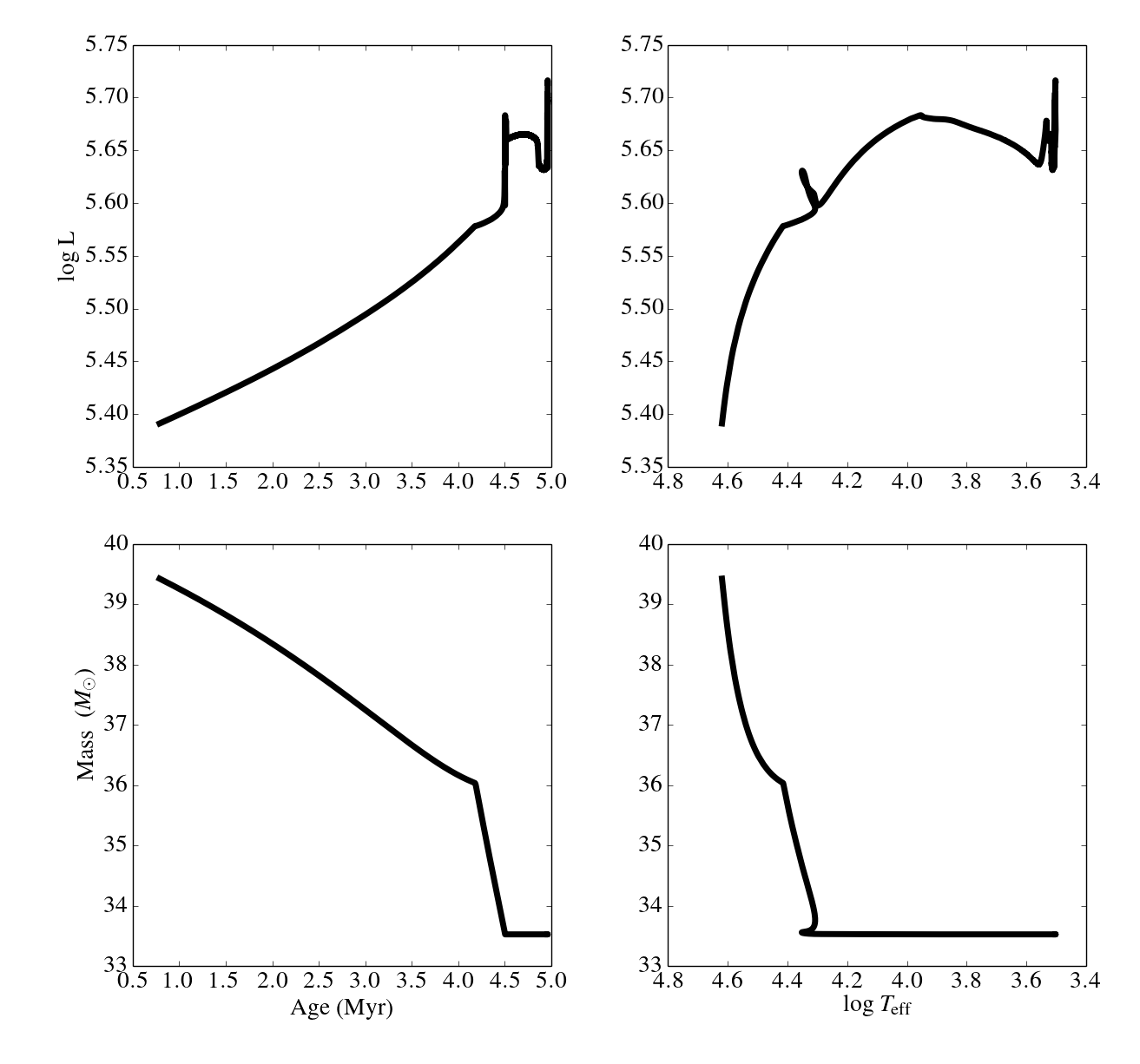}
\caption{\footnotesize Evolution of a 40\Mdot~ model with an initial rotation rate of 100\kms and \aov $=$ 0.1. This is shown for a variety of conventional plots such as the luminosity and mass as a function of stellar age (left upper and lower), as well as in a standard HRD (right upper), and finally mass as a function of effective temperature.}
\label{Quadrant}
\end{figure}

Eq.\,(\ref{ML}) shows that we can increase the luminosity by increased helium abundance. A minor helium enrichment in both the primary and secondary presents the possibility that the initial mass is not required to be insufficiently high to reach the dynamical mass. The observed helium enrichment corresponds to an increase in $L_{\rm init}$ by $\approx$ 30$\%$ or 0.11 dex. This offers a potential scenario that would allow for a higher luminosity and lower $M_{\rm init}$, nonetheless it is unlikely that the initial He abundance of HD\,166734 is enriched rather than having been exposed as fusion products at the surface during hydrogen burning. We consider this solution unlikely. 

Alternatively, the observed luminosity could be higher than would be required for the relevant initial mass due to the evolutionary phase at which these stars are currently undergoing. Beyond the TAMS, we observe an increased luminosity as models evolve to cooler temperatures. If HD\,166734 was in fact composed of helium burning objects, the observed luminosity could be explained by this increased post-TAMS. Yet when comparing our models with the observed T$_{\rm eff}$'s, we note that both objects remain too hot to be post-MS objects, regardless of \aov, thus excluding later evolutionary phases as a viable solution. 

We observe some models that reach the dynamical mass of the primary due to higher mass-loss rates relative to the higher initial mass, though even these models must be excluded due to the observed T$_{\rm eff}$ since the dynamical mass is only reached during the bi-stability regime at a much cooler temperature than observed. Scaling factors and dependencies between $\dot{M}$, \aov~ and \vsini~ present a complex situation to break into linear effects.

We constrain our models with HD\,166734 observations by utilising a variety of plots for consistency between mass, luminosity, temperature, and age; see figure \ref{Quadrant}. We explore the HRD position and compare this with the spectroscopic HRD (sHRD), which removes uncertainties with distance and luminosity. Simultaneously, we correlate ages of the primary and secondary with dynamical masses and mass-loss rates. Figure \ref{Quadrant} illustrates the relevant plots comparing HD\,166734 to reproduce the observed masses and luminosities concurrently.
 
\begin{figure}
\centering
\includegraphics[width = 8cm]{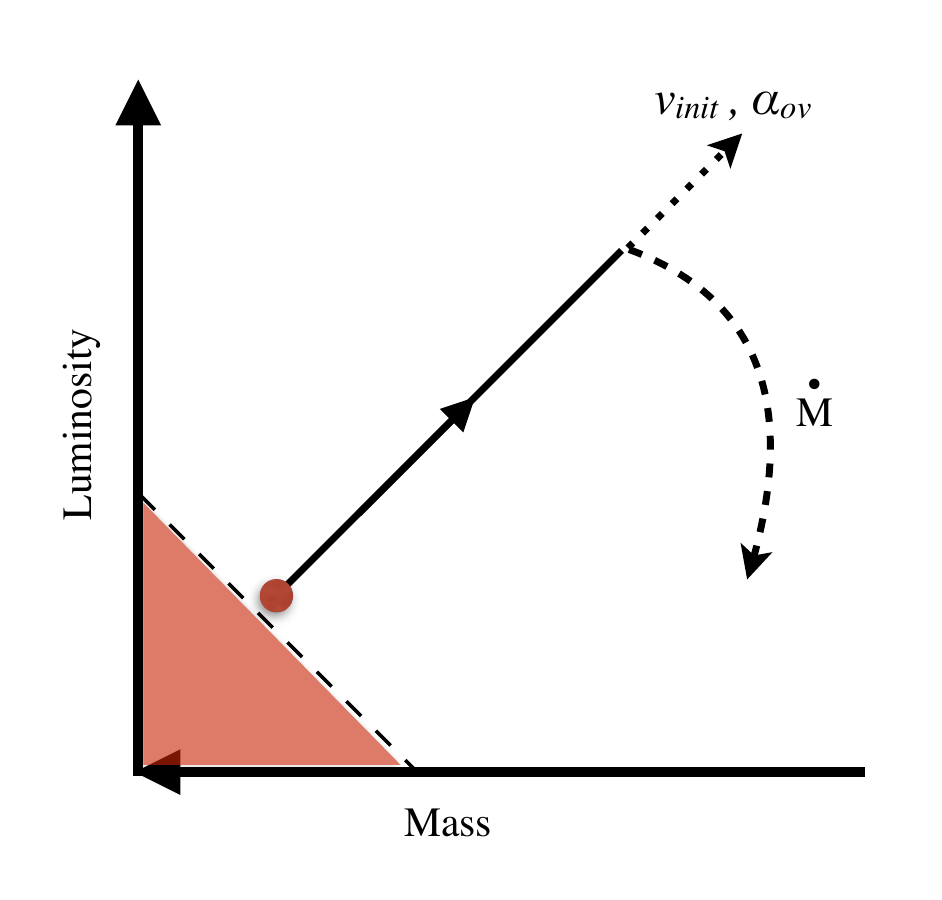}
\caption{\footnotesize Illustration of the Mass-Luminosity plane with a typical evolutionary track entering the ZAMS at the red dot, evolving along the black arrow. The dotted vector suggests how increased rotation and/or convective overshooting may extend the $M-L$ vector. The curved dashed line represents the gradient at which mass-loss rates affect this $M-L$ vector. The red solid region represents the boundary set by the mass-luminosity relationship, and as such is forbidden. }
\label{Cartoon}
\end{figure}

\cite{Maeder1986} discussed the complexity of mixing processes that apply to stellar evolution and the disentanglement required to understand the linear effects of each process fully. Mass loss is thought to dredge up fusion products to the surface while diminishing the core mass, extending the MS lifetime at the expense of the He-burning lifetime. In this respect, stellar winds behave similarly to convective overshooting or rotational mixing, even in extreme cases where a star may evolve quasi-chemically homogeneously due to extensive mixing. 

Earlier models which solely employ mass loss as the mixing process may present a simpler solution to understanding the full effects of this process. Although it has been stressed that to reproduce observations such as in the 34 open clusters from \cite{Mill81}, an extended MS was required leading to conclusions that overshooting is required as an additional mechanism of mixing. 

Similarly, in section \ref{3} we emphasised the necessity of rotational mixing in reproducing observed surface abundances. Therefore, since overshooting, mass loss and rotation have similar effects on the MS lifetime and appearance of CNO products at the stellar surface, a method of separating these processes must be developed. 

Challenges in reproducing masses and luminosities simultaneously remained while comparing the HRD and mass-age plots. It was consequently thought to be more insightful to compare our models by mass and luminosity directly. Interpreting behavioural characteristics of physical processes in this way has opened a diversity of information on luminosity and mass, as illustrated in Fig.\,\ref{Cartoon}. 

\begin{figure}
\centering
\includegraphics[width = 9cm]{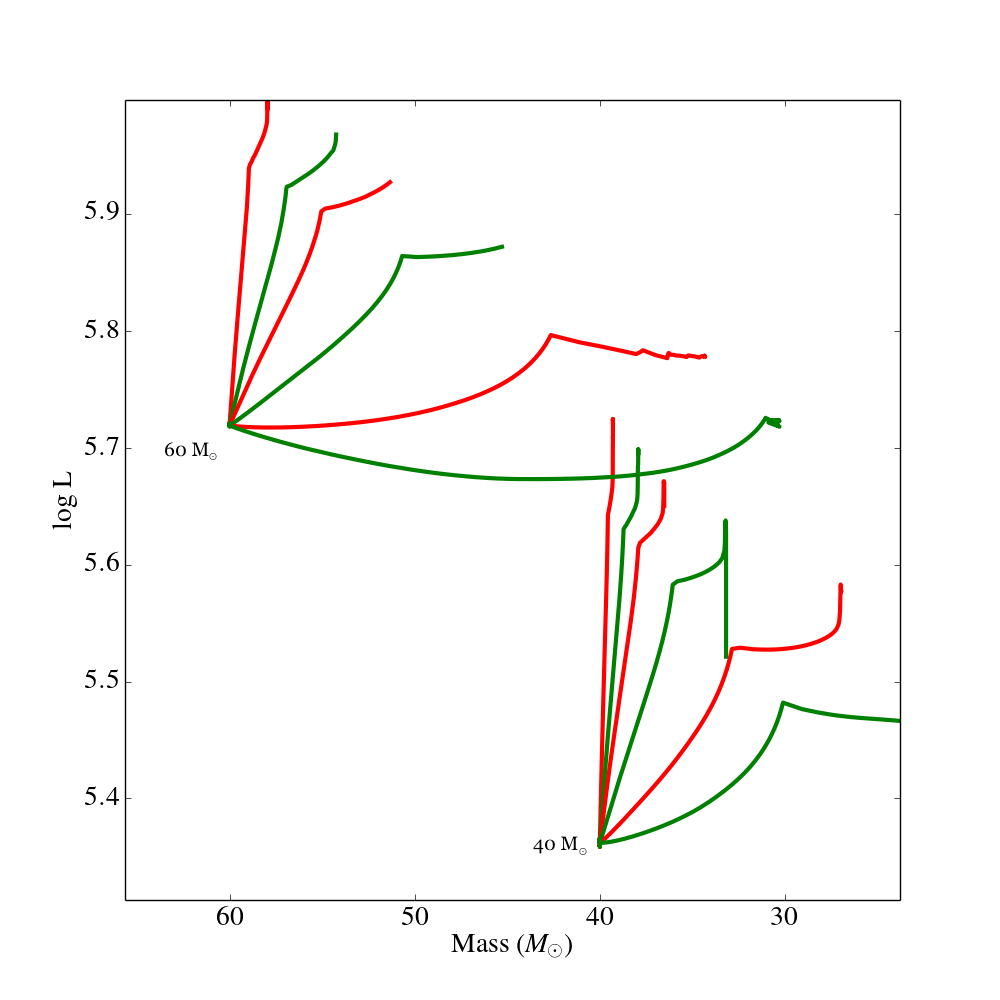}
\caption{\footnotesize Evolution of both 40\Mdot\, and 60\Mdot\, models with initial rotation rates of 100\kms and \aov $=$ 0.1 are shown for a variety of factors of \citet{Vink01} mass-loss rate (0.1 - 3 times), demonstrating the gradients for each model represented by the green and red solid lines in the $M-L$ plane.} 
\label{MLHD}
\end{figure}

Figure \ref{Cartoon} highlights the key features in the Mass-Luminosity Plane. As the star evolves with time, the vector of mass and luminosity increases in length, since MS stars increase in luminosity due to hydrogen burning. In this sense the $M-L$ plot is similar to the HRD in that time can be interchanged with temperature, since we also follow the vector length with respect to temperature, reaching characteristics such as the bi-stability jump. Figure \ref{Cartoon} demonstrates the evolution of a theoretical model to a particular age or temperature, by which we can compare this point with observations (e.g. an observed effective temperature).

\begin{figure}
\centering
\includegraphics[width = 9cm]{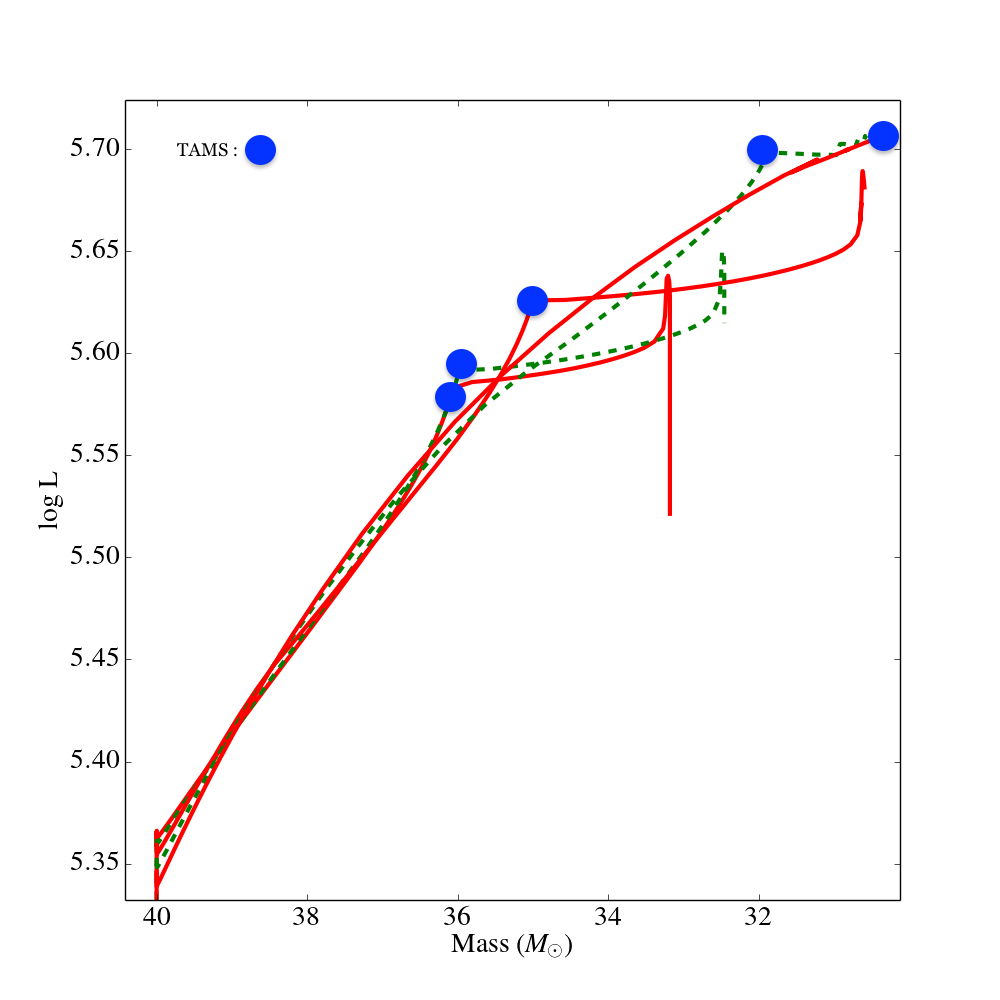}
\caption{\footnotesize Evolution of a 40\Mdot~ model with a factor of unity of the mass-loss prescription and \aov $=$ 0.1 shown for a variety of initial rotation rates from 100-500\kms. The length of the $M-L$ vector at a given evolutionary stage can be extended via increased rotation as shown by the blue dots corresponding to TAMS for each model.}
\label{MLrot}
\end{figure}

\begin{figure}
\centering
\includegraphics[width = 9cm]{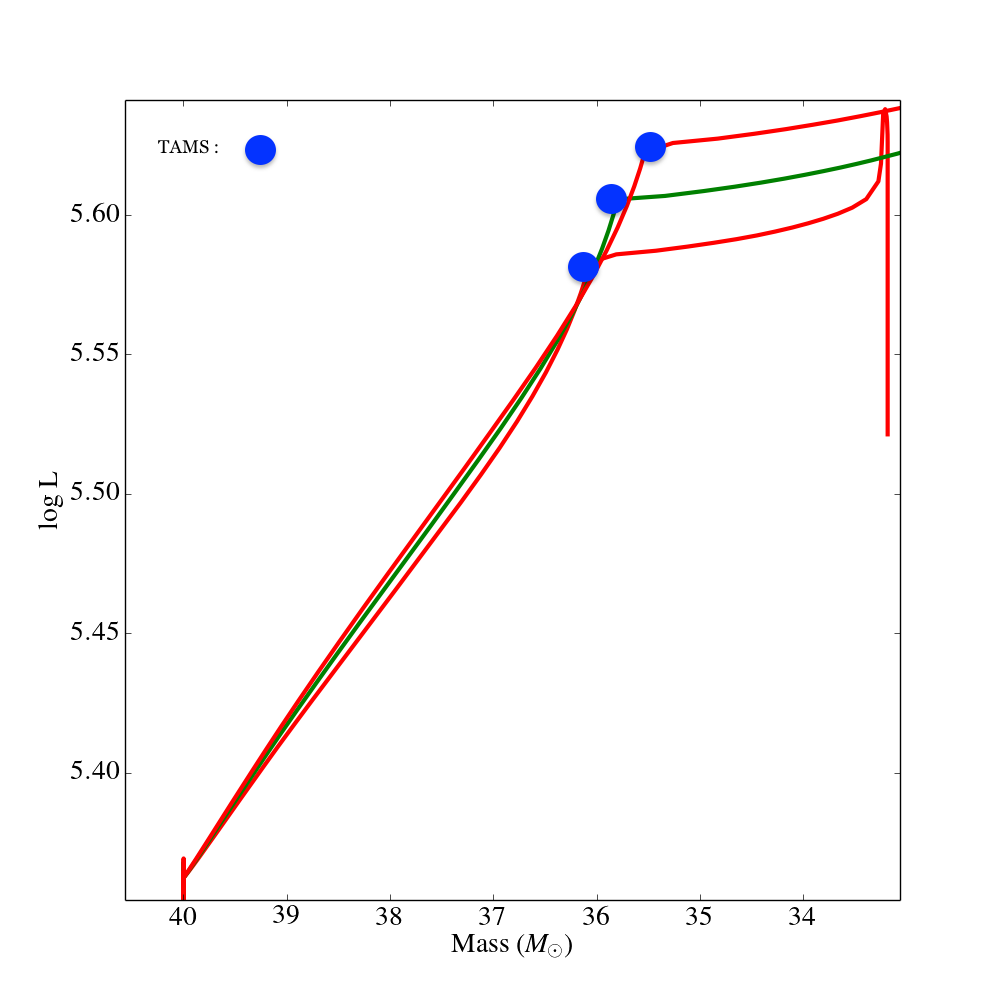}
\caption{\footnotesize Evolution of a 40\Mdot~ model with an initial rotation rate of 100\kms~ and a factor of unity for the mass-loss prescription shown for a variety of overshooting \aov $=$ 0.1, 0.3, 0.5. The blue dots correspond to TAMS for each model, demonstrating the increase in luminosity or decrease in mass for an increase in overshooting of \aov $=$ 0.1 - 0.5. This illustrates the possible further extension of a vector in this plane by extending \aov.}
\label{MLov}
\end{figure}

We note that the gradient of this vector is reliant on the mass-loss rate or in this case factors of the mass-loss prescription from \cite{Vink01}. Unsurprisingly, this feature becomes more prominent with higher masses, for example 60\Mdot\ compared to that of a 20\Mdot\ model, (see Fig. \ref{MLHD}). We find that the position of the vector at a given evolutionary phase or temperature can only be further extended in length by increased rotation or overshooting \aov, since greater mixing leads to higher luminosities (see Fig. \ref{MLrot} and \ref{MLov}), which have a higher mass-loss rate and subsequently a lower mass. 

When analysing our grid of models for the mass range 8-60\Mdot\ we found a set of features in the $M-L$ plane that provide fundamental boundaries to stellar evolutionary models. Figure \ref{Cartoon} illustrates one of these boundaries by a red solid forbidden region, by which the mass-luminosity relation (see Eq. \ref{ML}) sets the initial mass and luminosity. As a result of this relationship, stellar evolution models cannot lie within the red forbidden region. 

Similarly, if the length of the vector in the $M-L$ plane increases not only with time, but also temperature (as in the HRD) then we can adjust the length of our model based on an observed temperature. Thus we set an initial position and a final position in the $M-L$ plane for our models based on observed stellar parameters such as mass, luminosity and temperature. We can then utilise these positions to better understand processes such as rotation, mass loss and overshooting, since these all have an affect on our now "measured" vector length. 

Figures \ref{MLHD}, \ref{MLrot}, and \ref{MLov} each illustrate a process which influences the length or gradient of our vector in the $M-L$ plane. Figure \ref{MLHD} demonstrates that the mass-loss rate dictates a steep or shallow gradient, which again must be reached with the initial and final positions determined by the boundaries shown (i.e. the black line representing the forbidden region, and observations illustrating the final position). Figure \ref{MLrot} shows the possibility of extending the length of the vector by increasing the initial rotation rate, hence enhancing the luminosity. Finally, we can further extend the vector length by overshooting as represented in figure \ref{MLov} if rotation can be constrained through other methods such as \vsini~ and surface enrichments.

The range of explored factors of \citet{Vink01} mass-loss prescription can be seen in Fig.\, \ref{MLHD} for models with initial masses 40\Mdot\, and 60\Mdot. As we would expect, the factor of mass-loss rate has a much larger effect at 60\Mdot~ than the 40\Mdot. We find that due to the 'forbidden' region highlighted in Fig.\, \ref{Cartoon}, the gradients of models with two-three times the \citet{Vink01} prescription are much too shallow to reach observed initial luminosities of a 60\Mdot\, star for example. 
 
Fig.\, \ref{MLrot} illustrates an increase in luminosity by 0.1 dex for an increase in rotation of 200-400\kms. We find models with initial rotation rates of 100\kms and 200\kms are indistinguishable in the $M-L$ plane, although a notable increase in luminosity occurs above 200\kms. We use the TAMS as a reference point (blue dots) for each model demonstrating the effects of increased mixing by rotation or overshooting. 

\section{Observational constraints}\label{5}

\subsection{HD\,166734 parameter space}
To determine the initial parameters of the system HD\,166734, we computed a collection of models that adapt our methods from sections \ref{3} and \ref{4}, for a variety of initial masses, mass-loss rates, \aov\ and rotation rates. Due to constraining observations we have reproduced dynamical masses, luminosities, and surface nitrogen abundances based on a selection of parameters. 

Since there are multiple solutions to the current evolutionary stage, we present a parameter space in which the system can be reproduced within observational errors. This is necessary as following models with increased rotation or overshooting leads to higher luminosities. For example, models with higher mass-loss rates requires lower initial masses and thus lower rotation rates. 

We can reject extreme factors of \cite{Vink01} mass-loss rates due to the initial mass boundary in figure \ref{Cartoon}, such that we can reproduce the system with factors 0.5 - 1.5 of the \cite{Vink01} recipe. For initial masses of 55 - 60\Mdot\ for the primary and 42 - 47\Mdot\ for the secondary, we find a range of relevant overshooting parameters of 0.3 - 0.5 and 0.1 - 0.4 for the primary and secondary, respectively. We also stress that when calibrating our theoretical models, we ensure that the factor of mass-loss recipe \citep{Vink01} remains constant between the two objects to reach the most reliable solution.

When fixing the mass-loss prescription to a factor unity of \cite{Vink01}, we predict initial masses of 55\Mdot\ and 45\Mdot\ for the primary and secondary respectively. Initial rotation rates have been selected such that observed surface N abundances are reproduced with 250\kms and 120\kms\, for the primary and secondary, respectively. Having fixed the mass-loss rate and rotation rates of our models, we utilise the $M-L$ plane to measure the necessary overshooting required to reach the observed mass, luminosity and effective temperature of the primary and secondary. We discover greater values of \aov\ required to reproduce these stellar parameters with the primary adopting \aov = 0.3 $\pm$ 0.1 and the secondary requiring extra mixing of \aov = 0.5 $\pm$ 0.1 to reach the observed luminosity.

\begin{figure}
\centering
\includegraphics[width = 9cm]{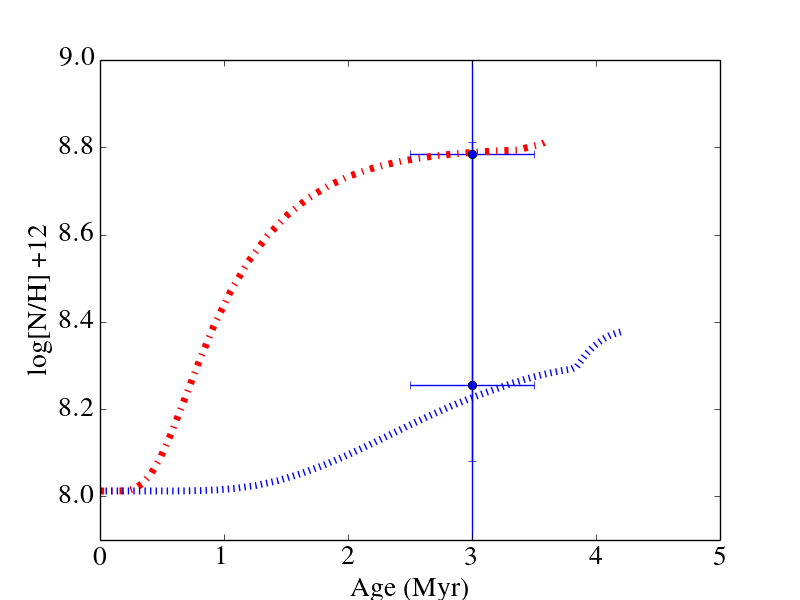}
\caption{\footnotesize Nitrogen enrichments from models in Fig \ref{HD166734}. Observations of HD\,166734 highlight the desired (3:1) ratio.}
\label{NHtdata}
\end{figure}

Figure \ref{HD166734} illustrates the evolution of the selected models which simultaneously reproduce observed luminosities and dynamical masses at an age of $\sim$3 Myrs. Observed nitrogen abundances are reproduced in figure \ref{NHtdata}, showing the observed 3:1 ratio of the primary to secondary. 

\begin{figure}
\centering
\includegraphics[width = 9cm]{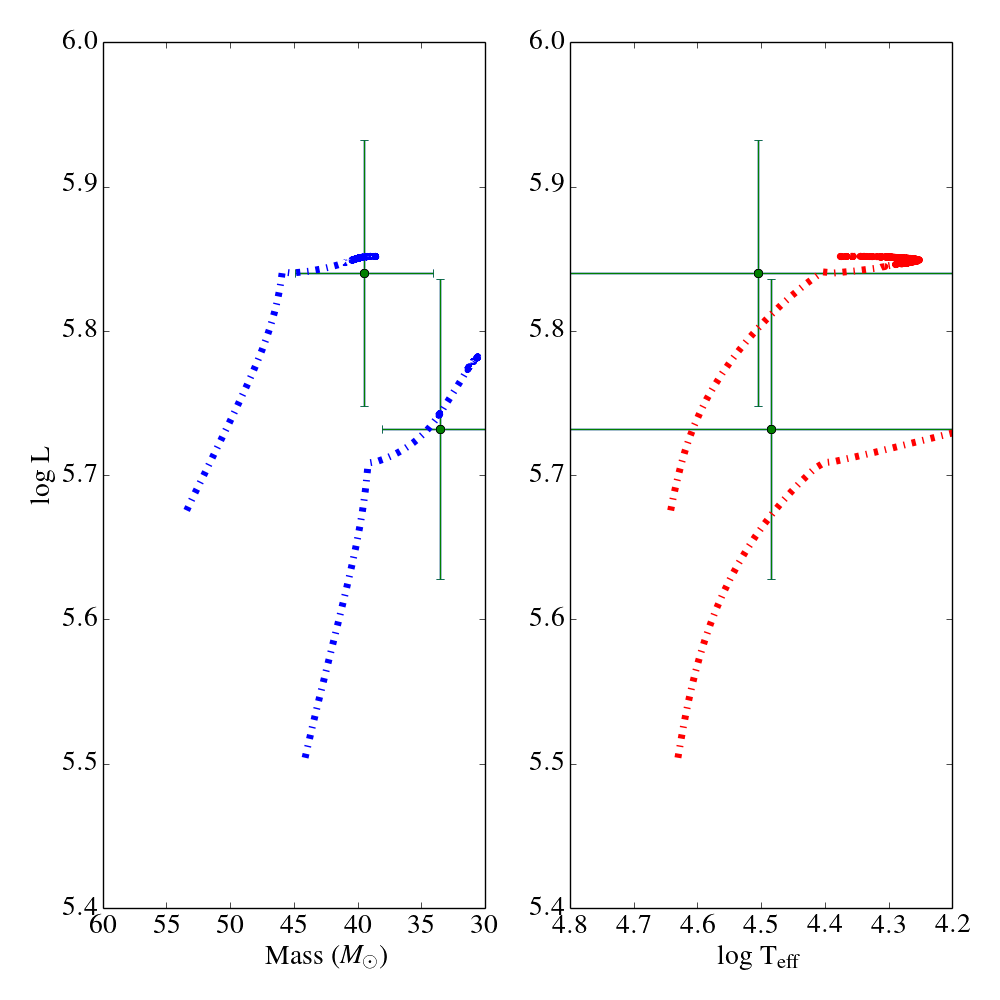}
\caption{\footnotesize HD\,166734 constrained models for primary and secondary in the $M-L$ plane (left) and HRD (right).}
\label{HD166734}
\end{figure}

\subsection{Applications from analysis}

We can further constrain the evolution of HD\,166734 due to constraints provided by \vsini~ and [N/H] abundances, after we constrain \aov\ in the $M-L$ plane. We seem to require a larger amount of core overshooting for the secondary star than for the more luminous primary. As the primary initial mass is of the order of 60\Mdot, effects such as envelope inflation \citep{GraefOwokVink, San15} and mass loss may potentially effect the stellar radius of the primary to a larger extent than it would for the secondary. Therefore, instead of arguing for an \textit{inverse} mass dependence of the \aov, we remain conservative, and consider the \aov\ determination of the secondary star as more secure than that of the primary.

\subsection{Galactic observational sample}

We aimed to consolidate our results from sections \ref{3}-\ref{5} by overlaying our calibration models for HD\, 166734 with a sample of 30 Galactic O stars from \cite{Markova} to ensure our calibration is representative of a larger sample, and not unique to our selected test-bed HD\,166734 only. 
The analysis by \cite{Markova} provided photospheric and wind parameters, including rotation rates, and surface N abundances by applying the model atmosphere code FASTWIND \citep{Puls05} to optical spectroscopy. Table \ref{sample} provides the key parameters we explored. 
We compared these Galactic data to our grid of models with the aim to constrain treatments of rotation and convection, and we contrast this with treatments from other evolutionary codes.

\begin{table}
\scriptsize
\caption{\label{t12}Galactic sample of O-stars.}
\centering
\begin{tabular}{lcccc}
\hline\hline
HD/CPD & $T_{\rm eff}$ [kK] & log(L/\Ldot) & $M_{\rm spec}$ [\Mdot] & $\rm [N/H]$ \\ 
\hline
HD 64568a & 48.0 $\pm$1.5 & 5.80 & 48.5 $\pm$17.9 & 8.18\\
HD 46223  & 43.5$\pm$1.5 & 5.58 &  38.9 $\pm$14.4 & 8.58\\
HD 93204a & 40.5$\pm$1.0  & 5.70 & 60.9 $\pm$22.5 & 7.78\\
CPD-59 2600a & 40.0 $\pm$1.0 & 5.40  & 40.3 $\pm$14.9  & 7.78\\
HD 93843a  & 39.0 $\pm$ 1.5 & 5.91 & 64.1$\pm$23.8 & 7.98\\
HD 91572a & 38.5 $\pm$1.0 & 5.35 & 32.7$\pm$12.1 & 8.37\\
HD 91824a & 39.0$\pm$1.0 & 5.37 & 32.7$\pm$12.1 & 8.48\\
HD 63005a & 38.5$\pm$1.0 & 5.52 & 34.4$\pm$12.7 & 8.58\\
CPD-58 2620a & 38.5$\pm$1.0 & 4.99 & 16.0$\pm$5.9 & 7.98\\
HD 93222 & 38.0$\pm$ 1.0 & 5.36 & 35.2$\pm$13.0 & 7.98\\
CD-47 4551 & 38.0$\pm$1.5 & 6.19 & 120.9$\pm$44.9 & 8.08\\
HD 94963a & 36.0$\pm$1.0 & 5.47 & 23.1$\pm$8.6 & 8.38\\
HD 94963b & & 5.62 & 32.4$\pm$12.0 & \\
HD 94370a & 36.0$\pm$1.0 & 5.36 & 29.9$\pm$11.1 & 7.78\\
HD 94370b & & 5.50 & 40.5$\pm$15.1 & \\
HD 92504 & 35.0$\pm$1.0 & 4.99 & 19.7$\pm$7.3 & 7.78\\
HD 75211 & 34.0$\pm$1.0 & 5.63 & 43.3$\pm$16.1 & 8.58\\
HD 46202 & 34.0$\pm$1.0 & 4.88 & 22.8$\pm$8.4 & 7.88\\
HD 152249 & 31.5$\pm$1.0 & 5.59 & 25.7$\pm$9.5 & 7.88\\
HD 151804 & 30.0$\pm$2.0 & 5.99 & 62.1$\pm$23.9 & 8.98\\
CD-44 4865 & 30.0$\pm$1.0 & 5.26 & 24.4$\pm$9.0 & 7.98\\
HD 152003 & 30.5$\pm$1.0 & 5.66 & 30.7$\pm$11.4 & 7.78\\
HD 75222 & 30.0$\pm$1.0 & 5.56 & 25.7$\pm$9.5 & 8.38\\
HD 75222a & & 5.67 & 32.8$\pm$12.2 & \\
HD 78344 & 30.0$\pm$1.0 & 5.60 & 33.3$\pm$12.3 & 8.58\\ 
HD 169582 & 37.0$\pm$1.0 & 6.10 & 86.1$\pm$32.1 & 8.98\\
CD-43 4690 & 37.0$\pm$1.0 & 5.53 & 29.5$\pm$10.9 & 8.38\\
HD 97848 & 36.5 $\pm$1.0 & 5.03 & 19.6$\pm$7.2 & 8.38\\
HD 69464 & 36.0$\pm$1.0 & 5.78 & 46.9$\pm$17.3 & 8.28\\
HD 302505 & 34.0$\pm$1.0 & 5.43 & 32.4$\pm$12.0 & 8.18\\
HD148546 & 31.0$\pm$1.0 & 5.70 & 35.7$\pm$13.2 & 8.98\\
HD 76968a & 31.0$\pm$1.0 & 5.58 & 29.8$\pm$1.0 & 8.18\\
HD 69106 & 30.0$\pm$1.0 & 5.09 & 21.8$\pm$8.1 & 8.00\\
\hline
\end{tabular}
\label{sample}
\tablefoot{
Sample of 30 O-type stars analysed by \cite{Markova}.
}
\end{table}

Over-plotting our models with the Galactic sample from \cite{Markova} not only allowed us to contrast evolutionary masses as derived from the spectroscopic HRD with the masses derived from the standard HRD, but also allowed us to compare our model grids with the prescriptions from \cite{Bonn11}. A sample of representative masses 20\Mdot, 40\Mdot\, and 60\Mdot\, of our grid, and the applied parameters from \cite{Bonn11} models (see Table \ref{BonnGEN}) for \aov and $\dot{M}_\text{rot, boost}$ that test our new prescriptions discussed in sections 3-4, are denoted in this work as Brott-like models. Note that we make comparisons based on two rotation rates for both Brott-like models and our new grid. 

\begin{figure}
\centering
\includegraphics[width = 9cm]{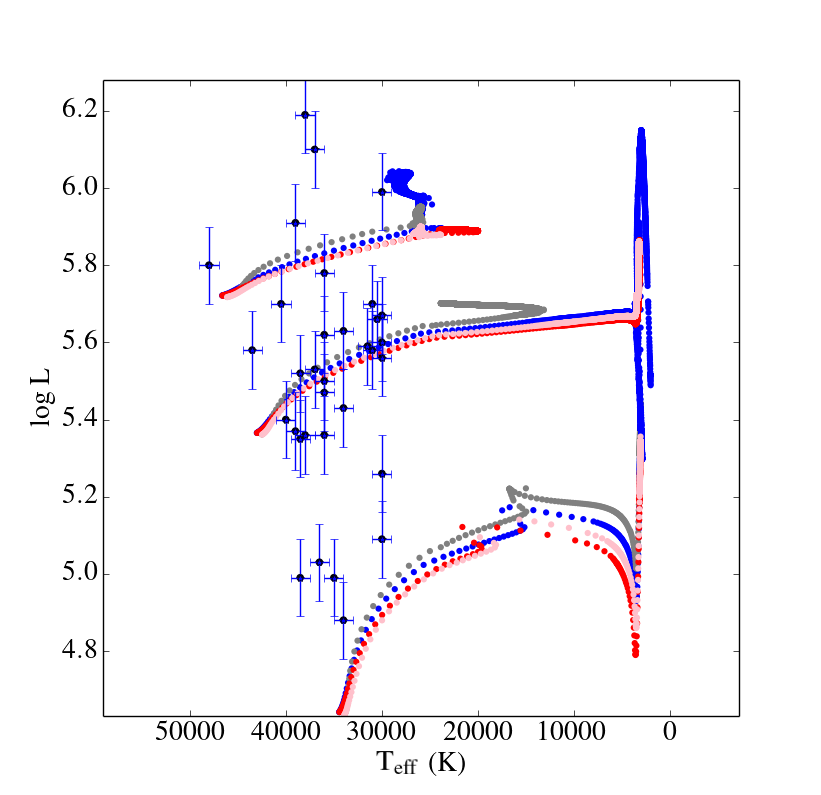}
\caption{\footnotesize Comparison of parameters taken from \cite{Bonn11} in red/pink with our new prescription in blue/grey, contrasted alongside our adopted Galactic sample of O stars from \cite{Markova}. }
\label{HRDBonn}
\end{figure}

\begin{figure}
\includegraphics[width = 9cm]{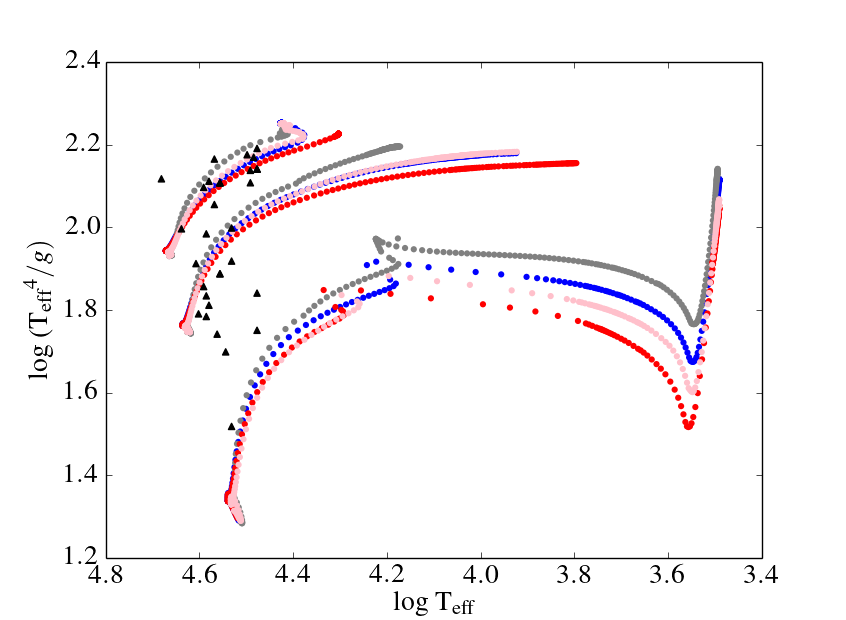}
\caption{\footnotesize Comparison of models as described in Fig. \ref{HRDBonn}. Data from \cite{Markova} are shown as black triangles in the form of a sHRD.}
\label{sHRDBonn}
\end{figure}

In figures \ref{HRDBonn} and \ref{sHRDBonn}, we present our tracks in blue and grey for 100\kms and 250\kms, respectively, and \cite{Bonn11} -like tracks in red and pink with 100\kms and 250\kms, respectively. We find a diminished discrepancy between $M_\text{evol, sHRD}$ $M_\text{evol, HRD}$ with our new models when compared to that of \cite{Bonn11} parameters by approximately 0.1dex as a result of increased luminosities with increased \aov, as well as the absence of the $\dot{M}_\text{rot, boost}$. This discrepancy is noted in \citet[pg. 12]{Markova} as a systematic difference in \cite{Gen12} models whereas \cite{Bonn11} models appear 10-20\% less massive in the HRD compared to the sHRD for masses above 30\Mdot. 

While exploring the possibility of a reduced discrepancy between evolutionary masses derived from the HRD and sHRD, we compared luminosities deduced by \cite{Markova} with the recent Gaia DR2 distance estimates \citep{GaiaDR2}. We found discrepancies in the distances of our sample and therefore luminosities when using the newly calculated distances through \cite{BailerJones}. However, when considering potential errors due to reddening, we found the reddening error to be larger than the already substantial error in the Gaia distances. Therefore, final answers would require a new spectroscopic analysis with proper consideration for reddening parameters and Gaia DR2 distances, which lies beyond the scope of this study.

\section{Grid analysis}\label{6}

Our systematic grid has been completed for two extreme values of \aov~ $=$ 0.1 and 0.5, since analysis from HD\,166734 invokes an argument for increased overshooting of \aov $=$ 0.5, and \aov $=$ 0.1 for allowing comparisons with previously published grids such as \cite{Gen12} models. For this reason, we show our tracks in figures \ref{HRDBonn} and \ref{sHRDBonn}, which are computed with our larger \aov~ prescription of \aov$=$ 0.5.
Table \ref{grid}. highlights the parameter space in which we compose our grid, compared with models from \citet{Gen12} and \citet{Bonn11} in table \ref{gridcomparison}. We identify the key variances as extra mixing by overshooting of up to \aov $=$ 0.5, and decreased mass loss by excluding rotationally induced mass loss. In order for our results to have relevance beyond the MS, we must ensure that observations of later evolutionary phases can be matched with our parameters. 

\begin{table*}[ht]
\caption{\label{gridcomparison}Comparison of parameter set-up with previously discussed evolutionary grids and this work.}
\centering
\begin{tabular}{l|ccc}
\hline\hline
Code & \cite{Bonn11} parameters & \cite{Gen12} parameters & This work : MESA\\
\hline 
$M_\text{initial}$ [\Mdot] & 5 - 60 & 8 - 120 &  8, 12, 16, 20, 25, 30, 35, 40, 45, 50, 55, 60 \\
$v_\text{initial}$ [\kms] & 0 - 550 & $v/v_\text{crit}$ $=$ 0.4 & 0, 100, 200, 300, 400, 500 \\
\aov & 0.335 & 0.1 & 0.1- 0.5  \\
Internal B-field & Spruit-Tayler & - & None\\
$\dot{M}_\text{rot, boost}$ factor & 0.43 & 0 & 0\\
\hline
\end{tabular}
\label{BonnGEN}
\end{table*}

\subsection{Red supergiant upper luminosity limit}
Red supergiants (RSGs) have been observed at luminosities up to log L/\Ldot $\approx$ 5.5 - 5.8 \citep{lev05, HD94}, with an observed cut-off after which RSGs are not created. In analysing our set of models, we compare final stages of evolution to RSG, BSG, or possible WR-type evolution (bluewards). We find that having fixed our mass-loss rates for clarity, overshooting has the dominant effect on the maximum mass/luminosities at which RSGs are formed. We note that the treatment of convection in the outer layer and the adopted mass-loss regime for this evolutionary phase also affects the position of the RSG.

If both \aov~ and $\dot{M}$ are fixed at lower values such as \aov$=$ 0.1, with standard \cite{Vink01} mass-loss rates, then RSGs are formed at luminosities of up to log L / \Ldot= 6.0, even for masses up to 60\Mdot. Since observations of RSGs suggest a lower cut-off in the range $\approx$5.5-5.8 dex, a higher value of \aov~ is required to match observations. Models which have adopted \aov$=$ 0.5 remain blue above log L/\Ldot $=$ 5.8 without evolving to RSGs, in agreement with the Humphreys-Davidson limit.

\begin{figure}[h]
\includegraphics[width = 10cm]{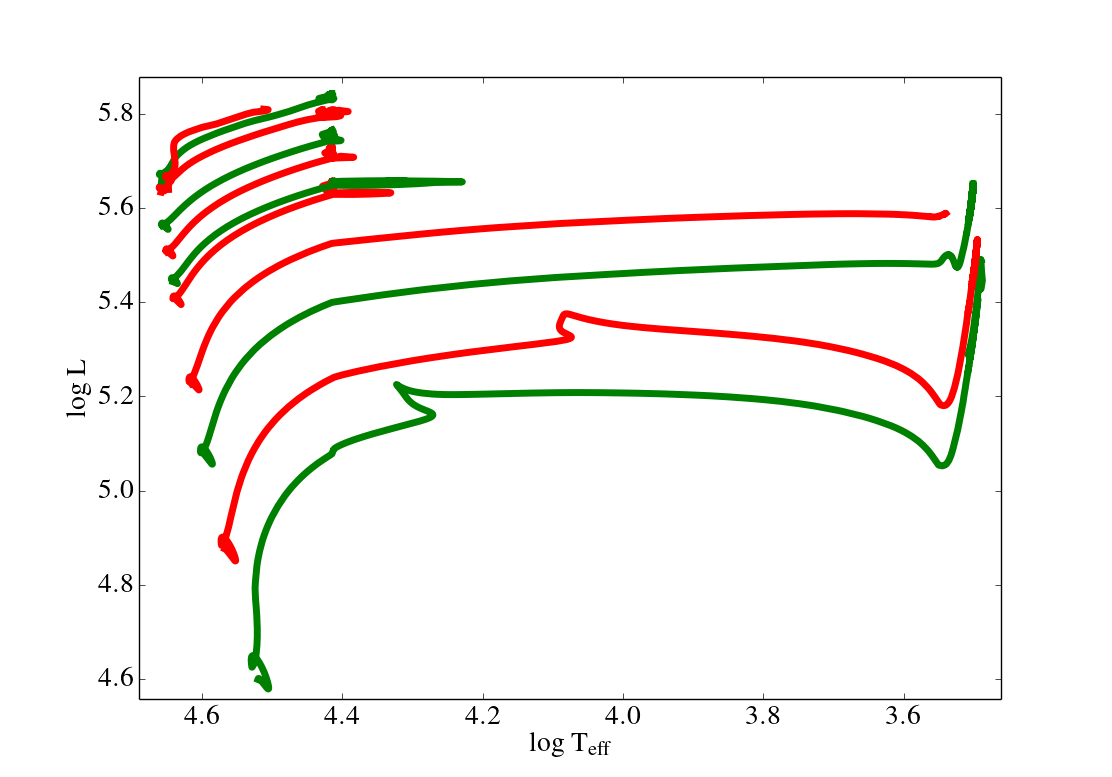}
\caption{\footnotesize Rotating models with \aov $=$ 0.5 and a factor of 1.5 $\dot{M}$, highlighting the post-MS evolution to the red for log L up to ~5.8, but above this limit remaining in the blue at TAMS. }
\label{RSG}
\end{figure}

We also examine final evolutionary phases for models from section \ref{4} with factors of $\dot{M}$ between 0.5 and 1.5, for \aov $=$ 0.1 and 0.5, finding that regardless of mass-loss rate (within our accepted parameter range), models which adopt \aov~ of 0.1 result in RSG evolution even at 60\Mdot. Figure \ref{RSG}. represents the observed luminosity cut-off for RSG evolution when implementing an enhanced overshooting of \aov $=$ 0.5.

\subsection{Compactness parameter}
Enhancing the overshooting parameter \aov\, has repercussions for the final fate of our models. The consequences of our grid results impact final mass estimates as well as compactness parameters which may dictate black hole and neutron star formation. We include our estimates of the compactness parameter  $\zeta_{2.5}$ for all final models, in which post-MS evolution is set as standard in MESA, as a function of \aov~ and initial rotation rate. \citet{oconnor} quantified the compactness of a presupernova stellar core as seen in Eq.\,(\ref{compacteq}), where $M = 2.5\Mdot~$ is selected as the relevant mass within which the iron core density gradient may be defined. The parameter $\zeta_{2.5}$ thus denotes how easily a presupernova stellar core explodes; with a low value leading to a more likely solution in which the star explodes rather than collapsing to form a black hole. \citet{SW14} found dependencies in the treatment of convection, including overshooting, with the explodability of presupernova models computed with MESA,

\begin{equation}
\centering
\zeta_{M} = \frac {M/{ \Mdot} }{R(M_{bary}=M)/1000km} .
\label{compacteq}
\end{equation}

\begin{figure}
\includegraphics[width = 10cm]{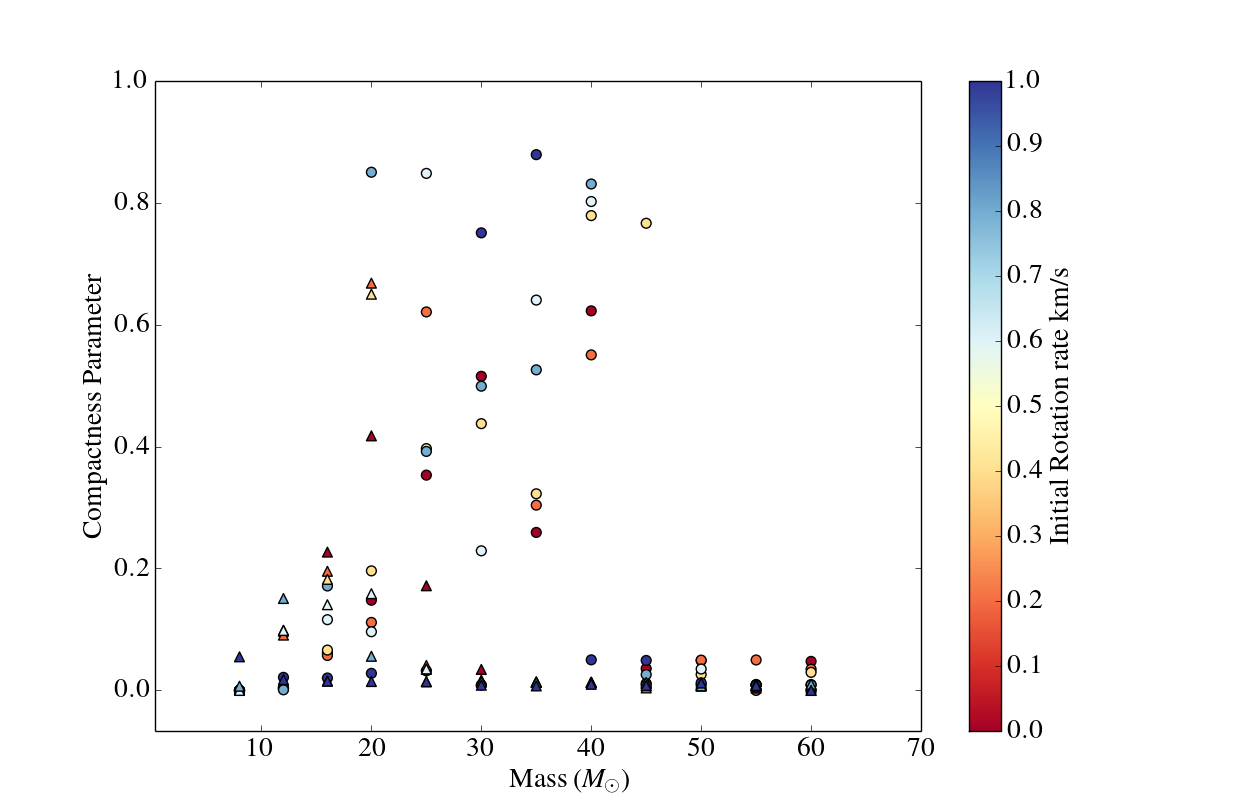}
\caption{\footnotesize Final value of compactness parameter for the initial mass range 20-60\Mdot ~ with \aov = 0.1 and 0.5, rotation rates of 0-500\kms. The circles represent models with \aov $=$ 0.1 and the triangles represent \aov $=$ 0.5. Rotation rates are shown as a fraction of the maximum 500\kms in the relevant colour bar.}
\label{compact}
\end{figure}

We note that with an extended convective core (\aov $=$ 0.5), and thus MS lifetime, it is more difficult to form black holes than for \aov $=$ 0.1 at the same mass range, as shown in Fig.\, \ref{compact}. A clear correlation with rotation rate is not reached, but can be compared via the representative colour bar. However, we note that rapidly rotating models with $v_{init}$ $=$ 500 \kms~ have a very low $\zeta_{2.5}$ and may explode more easily, regardless of \aov. 

\citet{Renzo17} presented values of $\zeta_{2.5}$ $\geq$ 0.25 for the mass range 15-30\Mdot with varying wind parameters, which are analogous to our results for a similar mass range with \aov $=$ 0.5. However we reach values of $\zeta_{2.5}$ $\geq$ 0.9 with \aov $=$ 0.1 for the mass range ~20-30\Mdot, suggesting this level of convective mixing may be less desirable for a high chance of explodability.

\section{Discussion and conclusions}\label{7}
We presented a calibrated grid of evolutionary models for the mass range 8 - 60 \Mdot, with initial rotation rates 0 - 500 \kms, and two values of overshooting \aov $=$ 0.1 and \aov $=$ 0.5 \citep{Higginsconference}. These models have been constrained based on results of our test-bed eclipsing binary HD\, 166734. 
We found that rotational mixing is necessary to reproduce the observed intermediate nitrogen enrichments, after first having explored the possibility that this could be achieved by overshooting and mass loss alone. In particular, we developed a method of reproducing the eclipsing binary HD\, 166734 based on the fundamental properties of mass and luminosity, utilising a new tool known as the mass-luminosity plane, the $M-L$ plane. This tool presents extensive information about the dominant physical processes for various mass ranges.

- First of all, the $M-L$ plane allows us to exclude very large increases or reductions in the standard mass-loss rates, via the gradient in the $M-L$ plane. More specifically, we can exclude mass-loss factors that lie beyond 0.5 - 1.5 times the \cite{Vink01} prescriptions. 

- Secondly, the extension of the data in the $M-L$ plane forces us to conclude that an additional process is required. Therefore, we favour large overshooting values of order \aov $=$ 0.5. The reproduced evolution of our test-bed high-mass binary HD\,166734 required this enhanced mixing by rotation and overshooting to increase the luminosity to that of the observed primary and secondary luminosities. 

 - Rotational mixing proves intrinsically necessary as the process whereby nitrogen is dredged to the surface in any intermediate quantity. Even though the process has been widely researched in the last few decades, the importance of reproducing observed surface abundances such as in HD\,166734 has not been sufficiently emphasised. We confirm that alternative mechanisms such as convection and mass loss cannot alone reproduce observed surface nitrogen abundances.

 - Finally, we disfavour the application of rotationally induced mass loss, in agreement with results from 2D computations of \cite{MullerVink}, as interacting processes artificially altering the initial mass-loss rate, leads to an entangled set of processes that cannot be separately constrained. The evolution of HD\,166734 cannot be reproduced with the inclusion of this theory.

If we compare observations to our new prescriptions of overshooting, mass loss, and rotation, we now open the possibility of an extended MS width, reinforcing the argument of B supergiants still being core H-burning objects \citep{Vink10}.

%\newpage
\bibliographystyle{aa} % style aa.bst
\bibliography{diff.bib} % your references 

\begin{acknowledgements} We thank Andreas Sander and the referee for constructive comments, and we acknowledge the MESA developers for facilitating the use of their publicly available code.
\end{acknowledgements}

\begin{appendix}
\section{Calibrated grid of models}\label{gridapp}
\begin{figure}[h]
\centering
\includegraphics[width = 9cm]{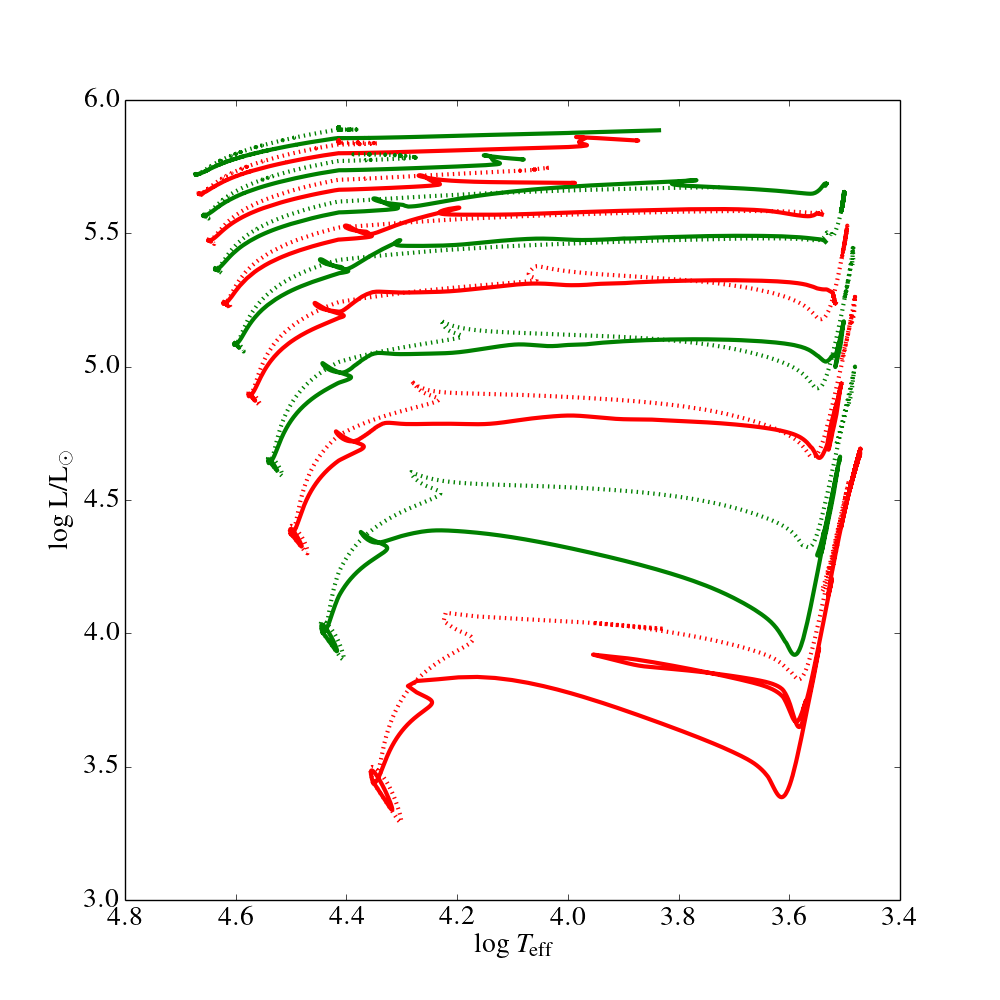}
\caption{\footnotesize Grid results for the non-rotating models in the mass range 8 - 60\Mdot employing \aov $=$ 0.1 (solid lines) and  \aov $=$ 0.5 (dashed lines).}
\label{grid0}
\end{figure}

\begin{figure}[h]
\centering
\includegraphics[width = 9cm]{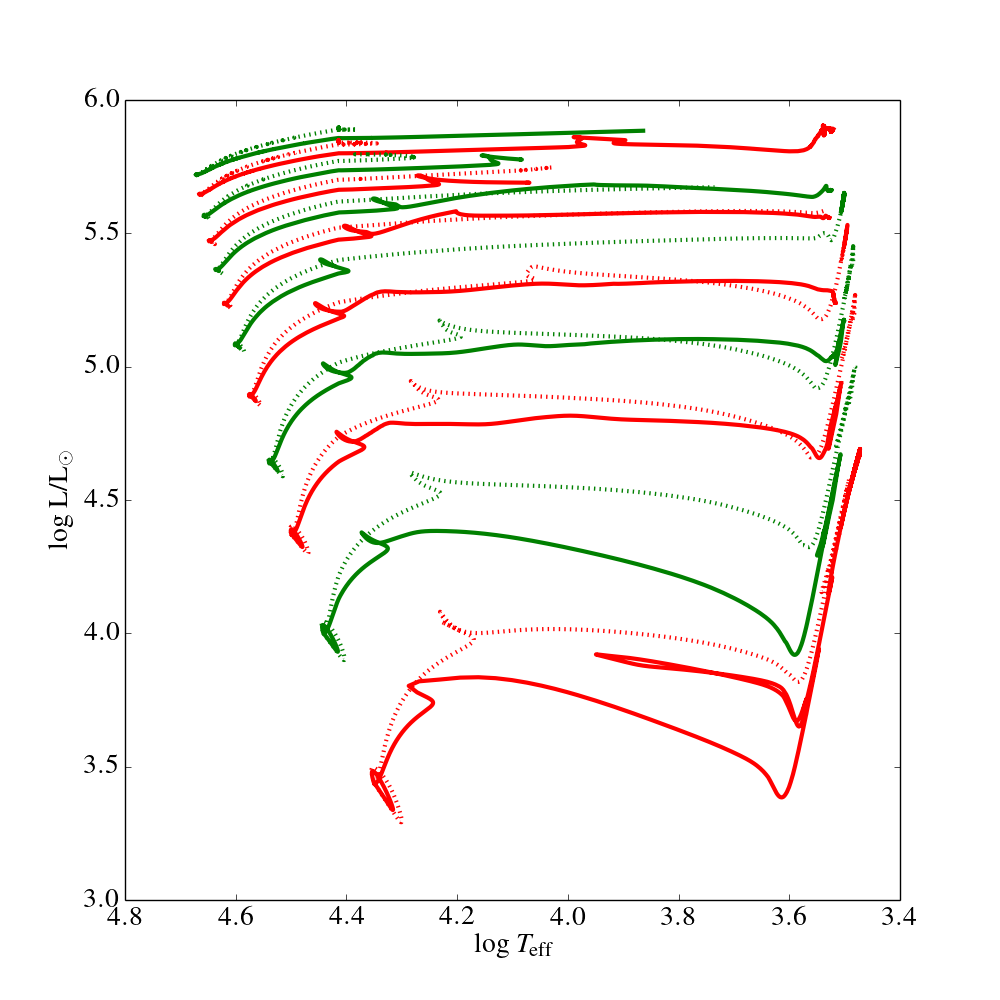}
\caption{\footnotesize Grid results for the rotating models with initial rotation rates of 100\kms for the mass range as mentioned in Fig. \ref{grid1},\aov $=$ 0.1 (solid lines) and  \aov $=$ 0.5 (dashed lines).}
\label{grid1}
\end{figure}

\begin{figure}[h]
\centering
\includegraphics[width = 9cm]{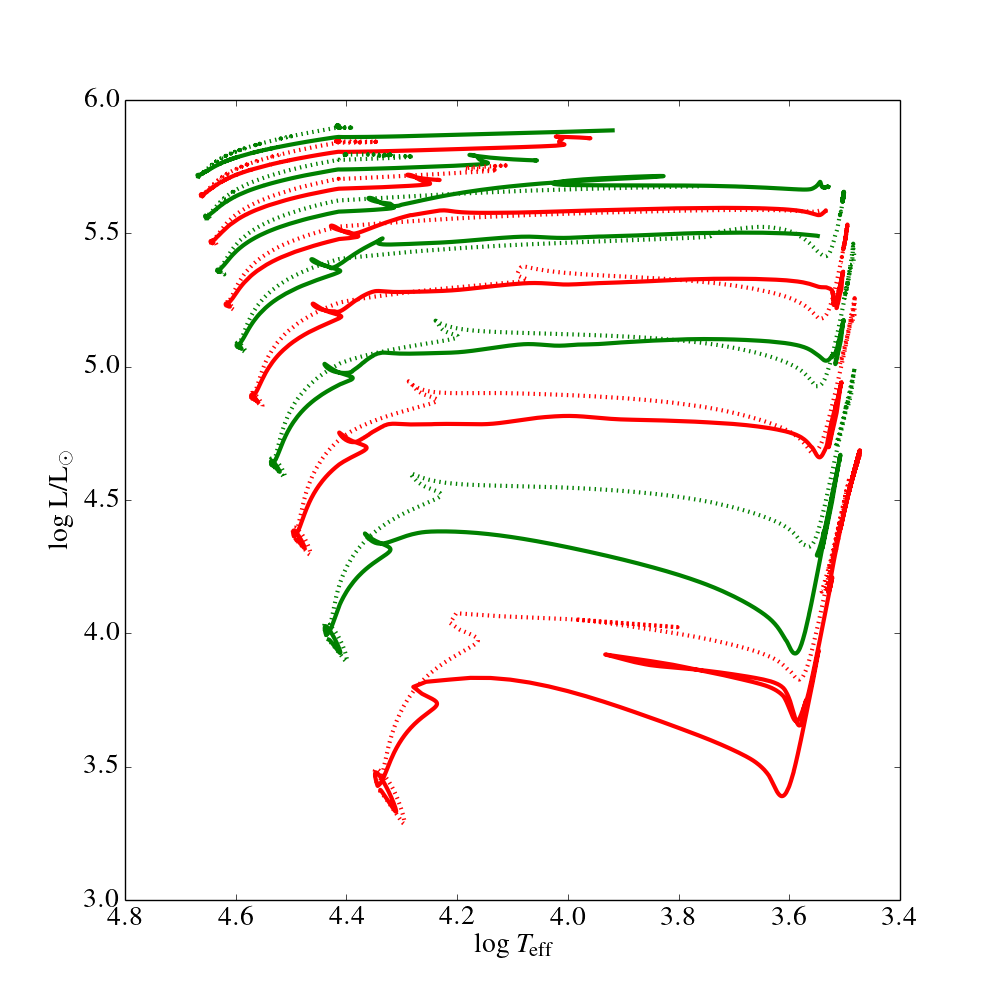}
\caption{\footnotesize Grid results for the rotating models with initial rotation rates of 200\kms for the mass range 8 - 60\Mdot employing \aov $=$ 0.1 (solid lines) and  \aov $=$ 0.5 (dashed lines).}
\label{grid2}
\end{figure}

\begin{figure}[h]
\centering
\includegraphics[width = 9cm]{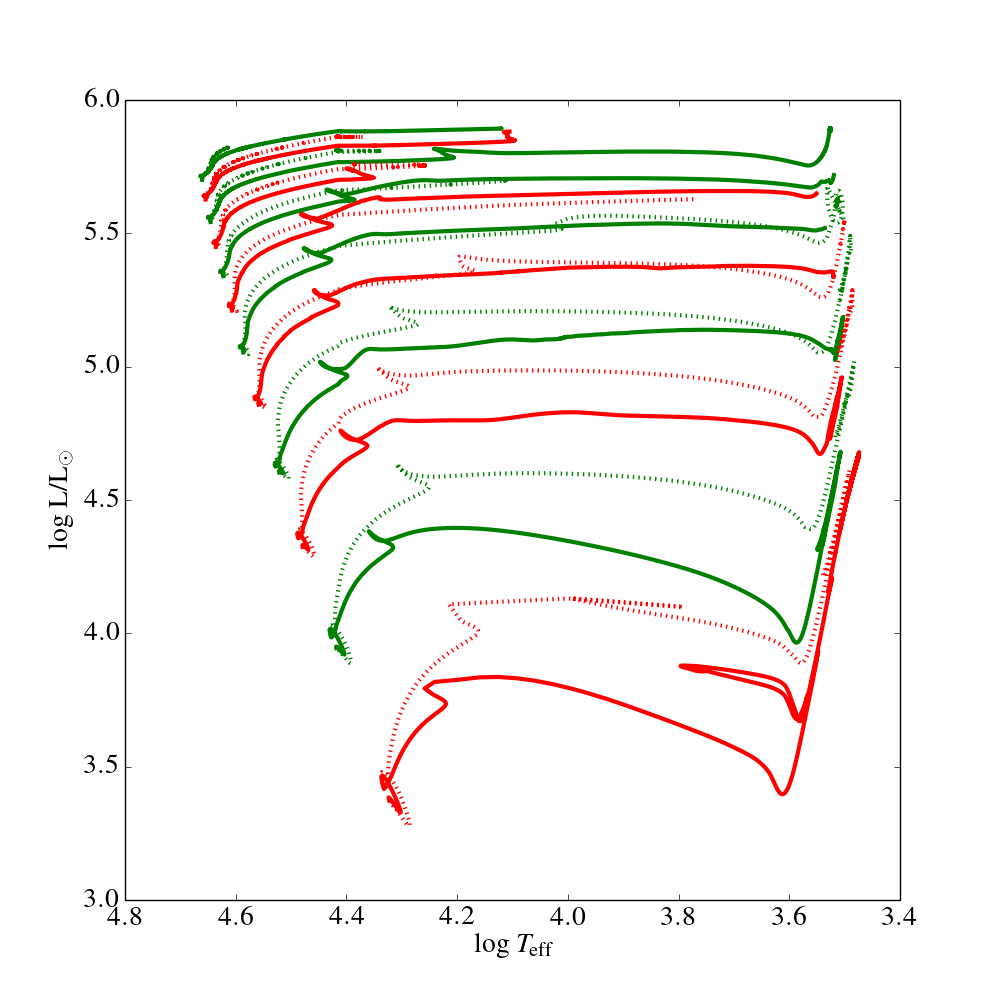}
\caption{\footnotesize Grid results for the rotating models with initial rotation rates of 300\kms for the mass range as mentioned in Fig. \ref{grid1}, \aov $=$ 0.1 (solid lines) and  \aov $=$ 0.5 (dashed lines).}
\label{grid3}
\end{figure}

\begin{figure}[h]
\centering
\includegraphics[width = 9cm]{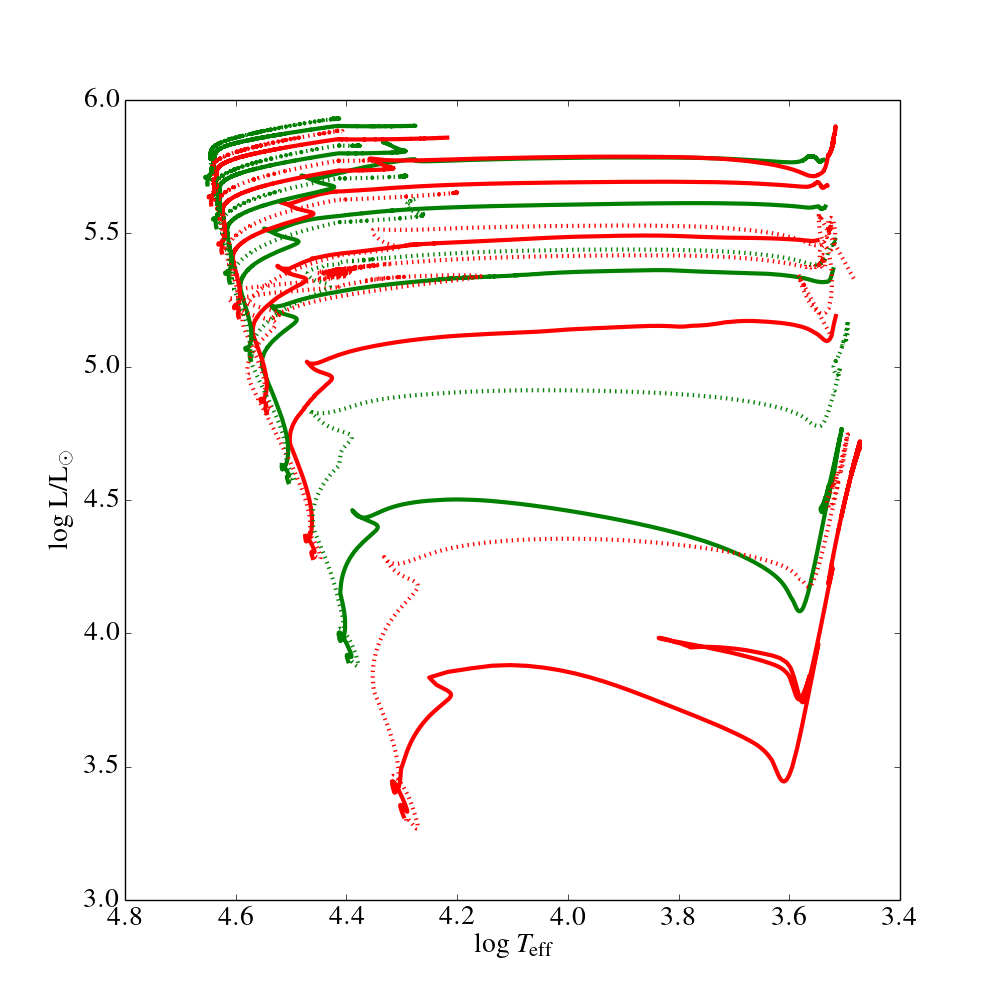}
\caption{\footnotesize Grid results for the rotating models with initial rotation rates of 400\kms for the mass range as mentioned in Fig. \ref{grid1}, \aov $=$ 0.1 (solid lines) and  \aov $=$ 0.5 (dashed lines).}
\label{grid4}
\end{figure}

\begin{figure}[h]
\centering
\includegraphics[width = 9cm]{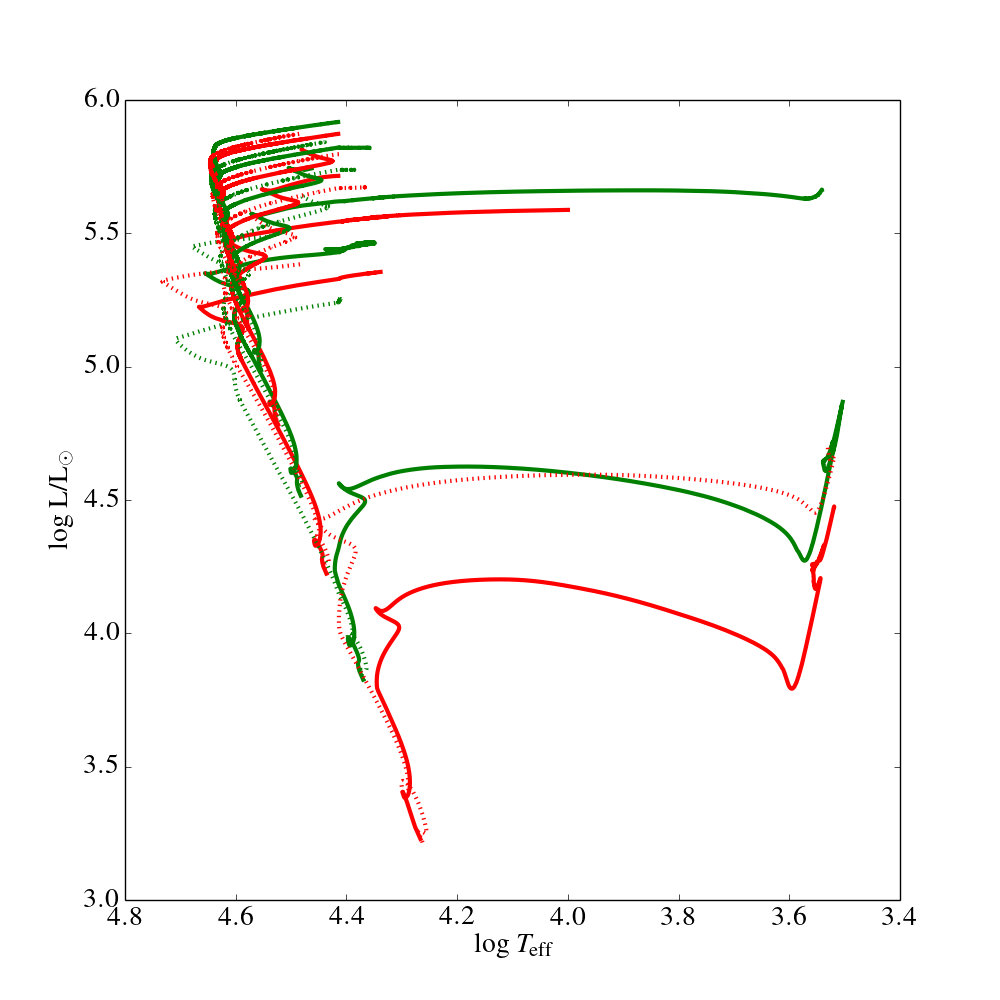}
\caption{\footnotesize Grid results for the rotating models with initial rotation rates of 500\kms for the mass range as mentioned in Fig. \ref{grid1}, \aov $=$ 0.1 (solid lines) and  \aov $=$ 0.5 (dashed lines).}
\label{grid5}
\end{figure}

\end{appendix}

\end{document}